\documentclass[fleqn,usenatbib]{mnras}

\usepackage{newtxtext,newtxmath}
\usepackage{siunitx}

\usepackage[T1]{fontenc}
\usepackage{xcolor}
\usepackage{multirow}
\usepackage{blindtext}
\usepackage{orcidlink}
\newcommand{\logrhk}[0]{\ensuremath{\log{R^{\prime}_\mathrm{HK}}}}

\newcommand{\edit}[1]{#1}
\DeclareRobustCommand{\VAN}[3]{#2}
\let\VANthebibliography\thebibliography
\def\thebibliography{\DeclareRobustCommand{\VAN}[3]{##3}\VANthebibliography}

\usepackage{graphicx}	
\usepackage{amsmath}	

\title[The TOI-5788 system]{Discovery and characterisation of two exoplanets orbiting the metal-poor, solar-type star TOI-5788 with TESS, CHEOPS, and HARPS-N}

\author[B. S. Lakeland et al.]{
Ben S. Lakeland\orcidlink{0000-0002-8122-2240},$^{1, 2}$\thanks{\href{mailto:b.s.lakeland@bham.ac.uk}{b.s.lakeland@bham.ac.uk}}
A. Mortier\orcidlink{0000-0001-7254-4363},$^{1}$\thanks{UKRI Future Leaders Fellow}
R. D. Haywood\orcidlink{0000-0001-9140-3574}, $^{2}$\thanks{STFC Ernest Rutherford Fellow}
S. Ulmer-Moll \orcidlink{0000-0003-2417-7006} $^{3, 4}$, 
Z. Garai \orcidlink{0000-0001-9483-2016}, $^{5, 6}$
\newauthor
A. Vanderburg, $^{45}$
J A. Egger \orcidlink{0000-0003-1628-4231}, $^{7, 8}$\thanks{ESA Research Fellow}
D. A. Turner\orcidlink{0009-0001-0867-9711},$^{1}$
D. Kubyshkina\orcidlink{0000-0001-9137-9818},$^{7, 9}$
A. C. M. Correia\orcidlink{0000-0002-8946-8579},$^{10}$
\newauthor
H. P. Osborn, $^{11, 12}$, 
L. A. Buchhave\orcidlink{0000-0003-1605-5666}, $^{13}$
L. Malavolta \orcidlink{0000-0002-6492-2085}, $^{14, 15}$
A. Bonfanti,$^9$
W. Boschin, $^{16}$
A. Cameron, $^{17}$
\newauthor
A. Castro-Gonz\'alez, $^{18}$
R. Cosentino, $^{16, 19}$
M. Damasso, $^{20}$
X. Dumusque, $^{18}$
D. Ehrenreich, $^{18, 21}$
Z. Essack, $^{22}$
\newauthor
S. Filomeno, $^{23, 24, 25}$
L. Fossati, $^{9}$
D. Gandolfi, $^{26}$
M. Gillon, $^{27}$
C. Hedges, $^{28}$
M. L\'opez-Morales, $^{29}$
\newauthor
G. Lacedelli, $^{30, 31}$
M. Lendl, $^{18}$
J. Maldonado, $^{34}$
G. Mantovan, $^{35, 14, 15}$
A. F. Martínez Fiorenzano, $^{16}$
\newauthor
P. F. L. Maxted, $^{36}$
C. Mordasini, $^{7, 11}$
B. Nicholson, $^{37}$
S. M. O'Brien, $^{38}$
L. Palethorpe, $^{39, 40}$
E. Palle, $^{30, 31}$
\newauthor
M. Pinamonti, $^{20}$
D. Rapetti\orcidlink{0000-0003-2196-6675}, $^{41, 42}$
I. Ribas, $^{43, 44}$
N. C. Santos, $^{32, 33}$
A. M. Silva, $^{32, 33}$
A. Sozzetti, $^{20}$
\newauthor
M. Stalport, $^{4, 27}$
G. Szab\'o, $^{6}$
S. Udry, $^{18}$
M. Vezie, $^{46}$
C. A. Watson, $^{38}$
and T. G. Wilson $^{47}$
\\
$^{1}$School of Physics and Astronomy, University of Birmingham, Edgbaston, Birmingham, B15 2TT, UK\\
$^{2}$Astrophysics Group, University of Exeter, Exeter, EX4 2QL, UK\\
$^{3}$Leiden Observatory, University of Leiden, Einsteinweg 55, 2333 CA Leiden, The Netherlands\\
$^{4}$Space sciences, Technologies and Astrophysics Research (STAR) Institute, Université de Li\`ege, Allée du 6 Août 19C, 4000 Liège, Belgium\\
$^{5}$Astronomical Institute, Slovak Academy of Sciences, 059 60 Tatranská Lomnica, Slovakia\\
$^{6}$ ELTE Gothard Astrophysical Observatory, Szent Imre h. u. 112, 9700 Szombathely, Hungary\\
$^{7}$Space Research and Planetary Sciences, Physics Institute, University of Bern, Gesellschaftsstrasse 6, 3012 Bern, Switzerland\\
$^{8}$European Space Agency (ESA), European Space Research and Technology Centre (ESTEC), Keplerlaan 1, 2201 AZ Noordwijk, The Netherlands\\
$^{9}$Space Research Institute, Austrian Academy of Sciences, Schmiedlstrasse 6, A-8042 Graz, Austria\\
$^{10}$CFisUC, Department of Physics, University of Coimbra, 3004-516 Coimbra, Portugal\\
$^{11}$Center for Space and Habitability, University of Bern, Gesellschaftsstra{\ss}e 6, 3012 Bern, Switzerland\\
$^{12}$ETH Zurich, Department of Physics, Wolfgang-Pauli-Stra{\ss}e 2, CH-8093 Zurich, Switzerland\\
$^{13}$DTU Space, Technical University of Denmark, Elektrovej 328, DK-2800 Kgs. Lyngby, Denmark\\
$^{14}$ Dipartimento di Fisica e Astronomia ``Galileo Galilei'', Università degli Studi di Padova, Vicolo dell'Osservatorio 3, Padova, IT-35122, Italy \\
$^{15}$ INAF - Osservatorio Astronomico di Padova, Vicolo dell'Osservatorio 5, Padova, IT-35122, Italy\\
$^{16}$ INAF - Fundación G. Galilei, Rambla J. A. Fernández Pérez 7, E-38712, Breña Baja (Islas Canarias), Spain\\
$^{17}$ Centre for Exoplanet Science / SUPA, School of Physics $\&$ Astronomy, University of St Andrews, North Haugh ST ANDREWS, Fife, KY16 9SS, UK\\
$^{18}$ Observatoire Astronomique de l’Université de Genève, Chemin Pegasi 51b, 1290 Versoix, Switzerland\\
$^{19}$ Osservatorio Astrofisico di Catania · Via Santa Sofia 78, 95123 Catania, Italy\\
$^{20}$ INAF - Osservatorio Astrofisico di Torino, Via Osservatorio 20, I-10025 Pino Torinese, Italy\\
$^{21}$ Centre Vie dans l’Univers, Faculté des sciences, Université de Genève, Quai Ernest-Ansermet 30, 1211 Genève 4, Switzerland\\
$^{22}$ Department of Physics and Astronomy, The University of New Mexico, 210 Yale Blvd NE, Albuquerque, NM 87106, USA\\
$^{23}$ INAF Osservatorio Astronomico di Roma, Via Frascati 33, I-00040 Monte Porzio Catone (RM), Italy \\
$^{24}$ Dipartimento di Fisica, Universit\`a di Roma Tor Vergata, Via della Ricerca Scientifica 1, I-00133 Roma, Italy\\ 
$^{25}$ Dipartimento di Fisica, Sapienza Università di Roma, Piazzale Aldo Moro 5, I-00185 Roma, Italy \\
$^{26}$ Dipartimento di Fisica, Universit\`a degli Studi di Torino, via Pietro Giuria 1, I-10125, Torino, Italy\\
$^{27}$ Astrobiology Research Unit, Universit\'e de Li\`ege, All\'ee du 6 ao\^ut 19, 4000 Li\`ege, Belgium\\
$^{28}$ NASA Goddard Space Flight Center, 8800 Greenbelt Rd, Greenbelt, MD 20771, USA\\
$^{29}$ Space Telescope Science Institute, 3700 San Martin Drive, Baltimore MD 21218, USA\\
$^{30}$ Instituto de Astrof\'{i}sica de Canarias (IAC), 38205 La Laguna, Tenerife, Spain\\
$^{31}$ Departamento de Astrof\'isica, Universidad de La Laguna (ULL), E-38206 La Laguna, Tenerife, Spain\\
$^{32}$ Instituto de Astrof'isica e Ciencias do Espa\c{c}o, Universidade do Porto, CAUP, Rua das Estrelas, 4150-762 Porto, Portugal\\
$^{33}$ Departamento de F'isica e Astronomia, Faculdade de Ci\^encias, Universidade do Porto, Rua do Campo Alegre, 4169-007 Porto, Portugal\\
$^{34}$ INAF – Osservatorio Astronomico di Palermo, Piazza del Parlamento 1, 90134 Palermo, Italy\\
$^{35}$ Centro di Ateneo di Studi e Attivit\`a Spaziali ``G. Colombo'' -- Universit\`a degli Studi di Padova, Via Venezia 15, IT-35131, Padova, Italy\\
$^{36}$ Astrophysics Group, Lennard Jones Building, Keele University, Staffordshire ST5 5BG, UK\\
$^{37}$ Centre for Astrophysics, University of Southern Queensland, West Street, Toowoomba, QLD, Australia\\
$^{38}$ Astrophysics Research Centre, School of Mathematics and Physics, Queen's University Belfast, Belfast, BT7 1NN, \\
$^{39}$ SUPA, Institute for Astronomy, University of Edinburgh, Royal Observatory, Blackford Hill, Edinburgh, EH9 3HJ, UK\\
$^{40}$ Centre for Exoplanet Science, University of Edinburgh, Edinburgh, EH9 3HJ, UK\\
$^{41}$ NASA Ames Research Center, Moffett Field, CA 94035, USA\\
$^{42}$ Research Institute for Advanced Computer Science, Universities Space Research Association, Washington, DC 20024, USA\\
$^{43}$ Institut de Ciencies de l'Espai (ICE, CSIC), Campus UAB, Can Magrans s/n, 08193 Bellaterra, Spain \\
$^{44}$ Institut d'Estudis Espacials de Catalunya (IEEC), 08860 Castelldefels (Barcelona), Spain\\
$^{45}$ Center for Astrophysics $\vert$ Harvard $\&$ Smithsonian, 60 Garden Street, Cambridge, MA 02138, USA\\
$^{46}$ Department of Physics and Kavli Institute for Astrophysics and Space Research, Massachusetts Institute of Technology, Cambridge, MA 02139, USA\\
$^{47}$ Department of Physics, University of Warwick, Gibbet Hill Road, Coventry CV4 7AL, United Kingdom
}

\date{Accepted XXX. Received YYY; in original form ZZZ}

\pubyear{2025}

\begin{document}
\label{firstpage}
\pagerange{\pageref{firstpage}--\pageref{lastpage}}

\maketitle

\begin{abstract}
We present the discovery and characterisation of two transiting exoplanets orbiting the metal-poor, solar-type star TOI-5788. 
From our analysis of six \textit{TESS} sectors and a dedicated \textit{CHEOPS} programme, we identify an inner planet (TOI-5788~b; $P =  6.340758\pm0.000030\,\si{\day}$) with radius $1.528\pm0.075\,\mathrm{R_\oplus}$ and an outer planet (TOI-5788~c; $P = 16.213362\pm0.000026\,\si{\day}$) with radius $2.272\pm0.039\,\mathrm{R_\oplus}$.
We obtained 125 radial-velocity spectra from HARPS-N and constrain the masses of TOI-5788~b and~c as $3.72\pm0.94\,\mathrm{M_\oplus}$ and  $6.4\pm1.2\,\mathrm{M_\oplus}$, respectively.
Although dynamical analyses indicate that a third planet could exist in a stable orbit between 8 and 14 days, we find no evidence of additional planets. 
Since the TOI-5788 system is one of the few systems with planets straddling the radius gap, and noting that there are even fewer such systems around metal poor stars, it is a promising system to constrain planet formation theories. 
We therefore model the interior structures of both planets. 
We find that TOI-5788~b is consistent with being a rocky planet with almost no envelope, or having an atmosphere of a high mean molecular weight. We find that TOI-5788~c is consistent with both gas-dwarf and water-world hypotheses of mini-Neptune formation.
We model the atmospheric evolution history of both planets. 
Whilst both scenarios are consistent with the atmospheric evolution of TOI-5788~c, the gas-dwarf model is marginally preferred.
The results of the atmospheric evolution analysis are not strongly dependent on stellar evolution.
This makes the system a promising target to test internal structure and atmospheric evolution models. 

\end{abstract}




\section{Introduction}

Despite the abundance and diversity of planetary systems unveiled since the discovery of the first exoplanets \citep{1992Natur.355..145W, 1995Natur.378..355M}, detailed understanding of planet formation and evolution pathways remains elusive.
One of the main legacies of the \textit{Kepler} mission \citep{Borucki} is the discovery of a bimodal distribution in the radii of small ($R\lesssim 4\,\mathrm{R_\oplus}$) exoplanets \citep{2017AJ....154..109F, 2023MNRAS.519.4056H}.
The formation and evolution pathways of these planets, between the size of Earth and Neptune, poses a particularly interesting problem. 
With no analogue in the Solar System, these planets are observed to form two distinct populations of \lq super-Earths\rq\ ($R\sim1.3\,\mathrm{R_\oplus}$) and \lq sub-Neptunes\rq\ ($R\sim2.5\,\mathrm{R_\oplus}$). 
These planets lie either side of the so-called \lq radius valley\rq.
Two main pathways are invoked to explain this bi-modality in exoplanet size. 
The first pathway requires both super-Earths and sub-Neptunes to form with a significant primordial H/He envelope. 
In this formalism, the super-Earths have lost almost all of their primordial atmosphere, with only the core remaining, whereas the larger mass of the sub-Neptunes enables them to retain their envelope. 
Within this model, two mechanisms are proposed to power the mass loss of the planets: high-energy (EUV or X-ray) flux from the host star \citep[photoevaporation; e.g.,][]{2013ApJ...775..105O, 2013ApJ...776....2L} or thermal emission from planet formation and bolometric stellar flux \citep[core-powered mass-loss; e.g.,][]{2019MNRAS.487...24G}.
We refer to this formalism as the \lq gas-dwarf\rq\ model.

The second formalism to explain the two \edit{distinct} populations of planets is the \lq water-world\rq\ model in which super-Earths and sub-Neptunes form with different planetary compositions \citep{2020A&A...643L...1V, 2022Sci...377.1211L}. 
This formalism holds that sub-Neptunes \edit{are mostly water-rich planets that formed outside the water ice line, where solid water ice can be accreted along with the silicates which are present throughout the disc}.
By contrast, super-Earths are expected to form from the silicate material present in the hotter region of the protoplanetary disc in which no solid volatile species can survive, \edit{producing the observed radius bimodality}.
A consequence of this \edit{model} is that sub-Neptunes observed within the ice line are expected to have migrated from the colder regions of the disc.
Whilst bulk measurements alone are often not sufficient to break the intrinsic degeneracies between these two formation pathways \citep{2010ApJ...712..974R}, in-depth studies of systems with well-characterised super-Earths and sub-Neptunes may reveal insights into the physics of their formation.

Another open question is on the link between planet and stellar chemical composition.
Since planets and stars form from the same nascent material, there is expected to be a link between the chemical composition of a planet and its host star. 
Recent studies of FGK stars with well-characterised planets have found a correlation between the iron mass fraction of a host star and rocky planets in its orbit \citep{Adibekyan21}. 
Studying planets orbiting metal-poor stars allows us to explore an even wider range of chemical compositions, as metal-poor stars are known to exhibit a wider range of alpha-enhancement, or [$\alpha/\mathrm{Fe}$],  \citep{2025PASA...42...51B}.
This is an important parameter space to explore as alpha elements are prominent in planetary mantles. 
Therefore any link between stellar and planetary compositions would manifest in variations in planetary makeup when studying systems across a wide range of $\left[ \alpha/\mathrm{Fe}\right]$ values.
To that end, well-characterised small planets around metal-poor stars are vitally important to understanding planet formation across diverse stellar systems.

In this paper, we present the discovery and characterisation of two planets orbiting the metal-poor solar type star TOI-5788. 
With both a super-Earth and sub-Neptune close to the inner edge of the radius valley, this system is ideal to study both the formation of super-Earths and sub-Neptunes and the link between planet compositions and the composition of their host star.
The remainder of this paper is structured as follows.
In Section~\ref{sec:data} we describe the data used in this study. 
In Section~\ref{sec:stellar_characterisation} we present the stellar analysis and derivation of stellar properties.
We present the transit and radial-velocity analyses in Sections~\ref{sec:transit_analysis} and~\ref{sec:rv_analysis}, respectively. 
From the adopted planetary parameters we obtain in these Sections, we model the interior structure and atmospheric evolution of both planets in Section~\ref{sec:characterisation}. 
We discus the potential for follow-up observations in Section~\ref{sec:follow_up} before concluding in Section~\ref{sec:conclusions}.

\section{Data} \label{sec:data}

\subsection{\textit{TESS} photometry} \label{sec:tess_data}

TOI-5788 (TIC 42883782) was observed by the Transiting Exoplanet Survey Satellite \citep[\textit{TESS}; ][]{2015JATIS...1a4003R} in Sectors 14, 40, 41, 54, 80, and 81.
The target was observed with 2-minute cadence in Sectors 14, 80, and 81; 10-minute cadence in Sectors 40, 41, and 54; and 20-second cadence in Sectors 80 and 81.
These data are made publicly available by the Mikulski Archive for Space Telescopes (MAST)\footnote{\url{https://archive.stsci.edu/missions-and-data/tess}}, and the image data were reduced and analysed by the Science Processing Operations Center \citep[SPOC;][]{jenkinsSPOC2016} at NASA Ames Research Center.

We modelled the instrumental systematics as a sum of moments of the spacecraft quaternion time series \citep[e.g., ][]{2019AAS...23332703V}, with long term modulations modelled via a basis spline.
The resulting light curve is similar to the original data reduction by the \textit{TESS} SPOC, although our data reduction benefits from retaining more of the higher-cadence data that is otherwise rejected by the SPOC reduction due to scattered light.

\begin{figure*}
    \centering
    \includegraphics[width=1\linewidth]{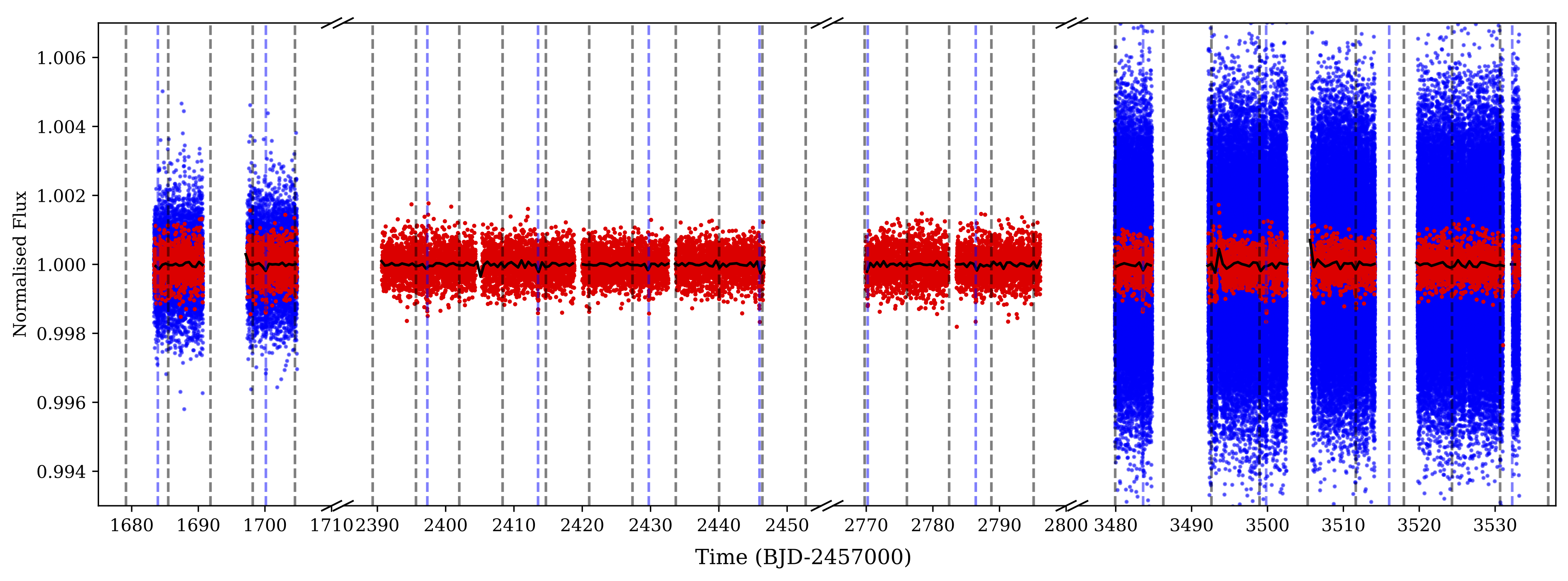}
    \caption{Six sectors of \textit{TESS} data for TOI-5788.
    The 2-minute cadence data of Sector 14 and 20-second cadence data of Sectors 80 and 81 are shown in blue.
    Sectors 40, 41, and 54 have a cadence of 10 minutes (red). 
We bin the light curves to 4.8 hours (black).
For Sectors with shorter cadence than 10 minutes, we show in red the 10-minute-binned data.
We show the times of mid transit for TOI-5788~b (grey dashed lines) and TOI-5788~c (blue dashed lines).}
    \label{fig:lightcurve}
\end{figure*}

From the first four sectors of \textit{TESS} data, two transit signatures were detected in a search of Full Frame Image (FFI) data by the Quick Look Pipeline (QLP) at MIT \citep{2020RNAAS...4..204H, 2020RNAAS...4..206H}. 
The TESS Science Office (TSO) reviewed the vetting information and issued an alert on 22 September 2022 \citep{2021ApJS..254...39G}.
The automatically derived orbital periods were $6.3346129 \pm 0.001121\,\si{\day}$ and $16.2134274\pm 0.0001179\,\si{\day}$. 
Fig.~\ref{fig:lightcurve} shows the six sectors of \textit{TESS} data.

\subsection{\textit{CHEOPS} photometry} \label{sec:cheops_data}

TOI-5788 was observed as part of a \textit{CHEOPS}  \citep{2021ExA....51..109B, 2024A&A...687A.302F} Guaranteed Time Observations programme (GTO; PI: Ulmer-Moll\edit{, Program number: 84, Program name: CompoSubNeptu.}). Three visits were planned for the transits of each planet, amounting to 39 \textit{CHEOPS} orbits between BJD 2460110 and 2460491.
Owing to a slightly incorrect value of the inner planet's orbital period reported on ExoFOP (see Section~\ref{sec:tess_data}), the scheduled \textit{CHEOPS} observations did not occur in transit for that planet. 
We therefore disregard the data from the three \textit{CHEOPS} visits that do not occur in transit, and retain only the transits of the outer planet.
\textit{CHEOPS} successfully observed the outer planet for all three planned transits.
We provide a log of the CHEOPS observations in Table~\ref{tab:cheops_log}.

\textit{CHEOPS} light curves were produced via the \textsc{PIPE} extraction technique.
Each \textit{CHEOPS} light curve was detrended against effects of spacecraft roll angle using the \textsc{pycheops} package \citep{Maxted22}.
The detrending vectors were chosen as the combination which minimises the Akaike Information Criterion \citep[AIC;][]{1974ITAC...19..716A} for each light curve. 
The detrending vectors were fit simultaneously with a transit model for each \textit{CHEOPS} light curve to obtain the optimally detrended light curve. 
The individual detrended light curves, with the transits retained in the data, were stitched and retained for later analysis.

\begin{table*}
    \centering
    \caption{Log of \textit{CHEOPS} observations. 
    Detrending vectors are listed for the transits of TOI-5788~c. 
    An incorrect ephemeris meant that all transits of TOI-5788~b were missed. 
    We therefore do not use these \textit{CHEOPS} data and do not list the relevant detrending vectors.
    }
    \begin{tabular}{ccccc}
    \hline \hline
         File Key&  Start date&  Duration&  Efficiency&  Planet\\
         & [UTC] & [h] &[\%]& \\\hline
         CH\_PR140084\_TG003502\_V0300 & 2023-07-17 10:59:39 & 10.3 & 67.0 & b \\
         CH\_PR140084\_TG003702\_V0300 & 2023-06-22 03:01:40 & 10.6 & 63.1 & b\\
         CH\_PR140084\_TG004101\_V0300 & 2024-06-29 20:17:19 & 9.3 & 72.1 & b  \\
         \hline
         CH\_PR140084\_TG003501\_V0300 & 2023-06-14 23:40:40 & 10.5 & 65.5 & c \\
         CH\_PR140084\_TG003701\_V0300 & 2023-06-15 17:28:40 & 9.8 & 67.5 & c  \\
         CH\_PR140084\_TG003503\_V0300 & 2023-08-02 15:52:40 & 10.8 & 62.8 & c \\
         \hline \hline
    \end{tabular}
    \label{tab:cheops_log}
\end{table*}

\subsection{HARPS-N spectroscopy} \label{sec:rv_data}

Through the High Accuracy Radial velocity Planet Searcher for the Northern hemisphere \citep[HARPS-N;][]{2012SPIE.8446E..1VC, 2014SPIE.9147E..8CC} Collaboration time (GTO until 2023A; A48TAC\_59, PI: Malavolta, thereafter), we obtained 125 high-resolution spectra of TOI-5788 for radial velocity (RV) follow up of these candidate planets. 
The HARPS-N spectrograph is installed on the 3.6~m Telescopio Nazionale Galileo (TNG) at the Observatorio del Roque de Los Muchachos on La Palma, Spain.
The design for HARPS-N is similar to its predecessor, HARPS \citep[installed on the ESO 3.6~m  telescope, ][]{2003Msngr.114...20M}; it covers a wavelength range of 383\,\si{\nano\metre} to 691\,\si{\nano\metre} and has an average resolving power of $115000$.

Observations were taken between 2022-Nov-06 and 2024-Nov-28.
The majority of the observations had exposure times of 900\,\si{\second}, but the final 32 observations (from 2023-Oct-25 onwards) were taken with exposure times of 1800\,\si{\second} in an effort to improve the signal-to-noise ratio (SNR).

HARPS-N spectra were reduced using version 3.0.1 of the HARPS-N Data Reduction Software
\citep[DRS;][]{2025arXiv251027635D},
giving RV and activity indicator measurements.
This version is similar to the previous 2.3.5 version of the DRS but with improved long-term stability.
RVs were extracted for TOI-5788 using a G8-type stellar template.
The final RV time series has a root-mean-square (RMS) scatter of 3.66\,\si{\metre\per\second}, and shows no evidence of long term trends. 
Along with RV measurements, the DRS calculates proxies for stellar activity such as the $S$-index to measure chromospheric emission,  full width at half maximum (FWHM) of the cross-correlation function (CCF) and the CCF bisector inverse slope (BIS).
We also follow the approach of \citet{1984ApJ...279..763N} and calculate the chromospheric emission ratio, \logrhk, from the S-index.

From the DRS reduction, there were five observations which displayed anomalously low $S$-index values. 
These $S$-index values, and their derived \logrhk measurements, are omitted from all analyses in this paper.
Since no other activity indicator, or the RVs themselves, showed anomalous values, we retain these measurements and only exclude the $S$-index measurements from our analyses.

\begin{figure}
    \centering
    \includegraphics[width=1\linewidth]{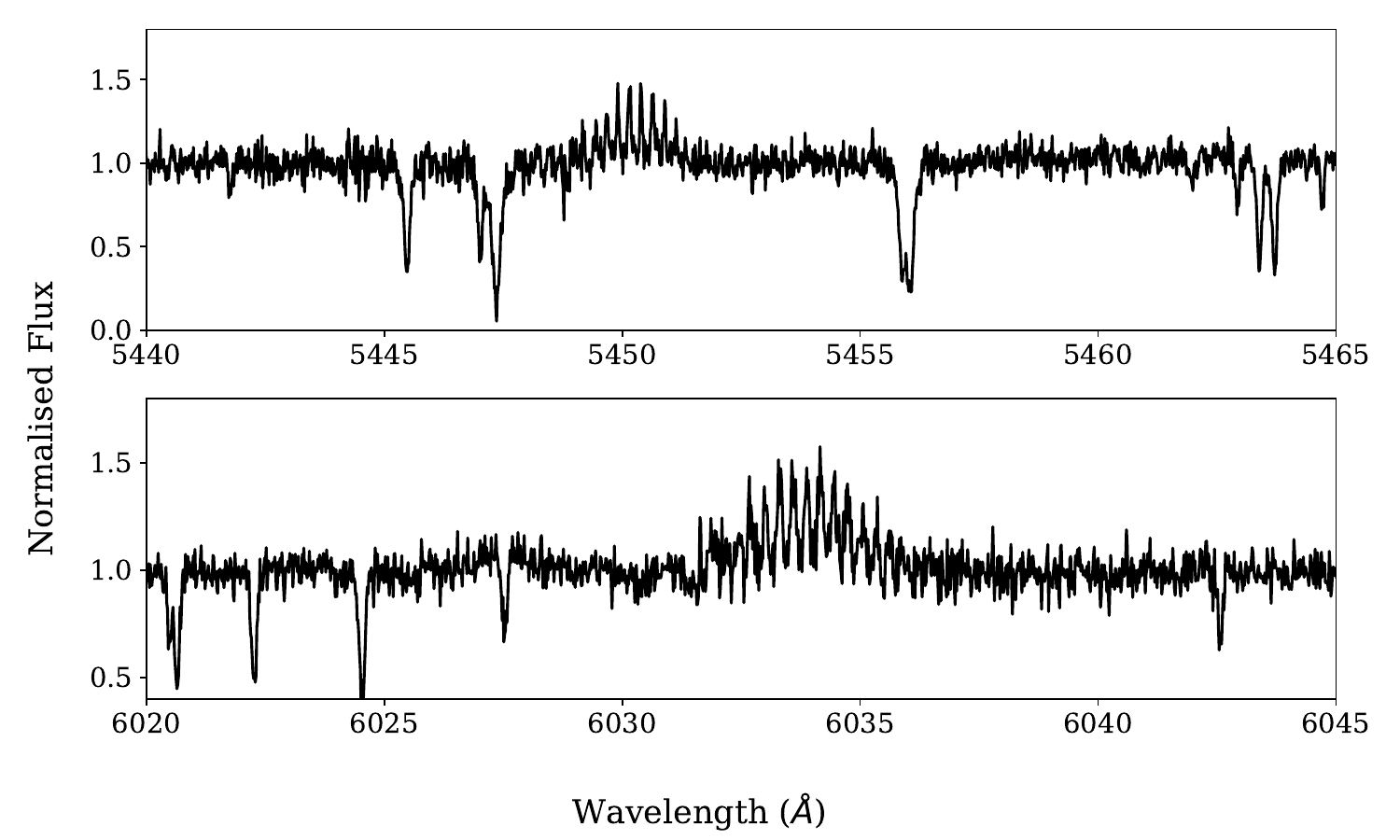}
    \caption{Fabry--P\'erot contamination of an arbitrarily selected HARPS-N spectrum of TOI-5788.
    The contamination manifests as regularly spaced emission lines, shown here at around 5450 and 6034~\AA.}
    \label{fig:fp_contamination}
\end{figure}

Furthermore, a number of spectra suffer from contamination from the Fabry--P\'erot etalon used to derive the wavelength solution. 
This manifests in spectra as anomalous emission lines, shown in Fig.~\ref{fig:fp_contamination}.

\begin{figure*}
    \centering
    \includegraphics[width=1\linewidth]{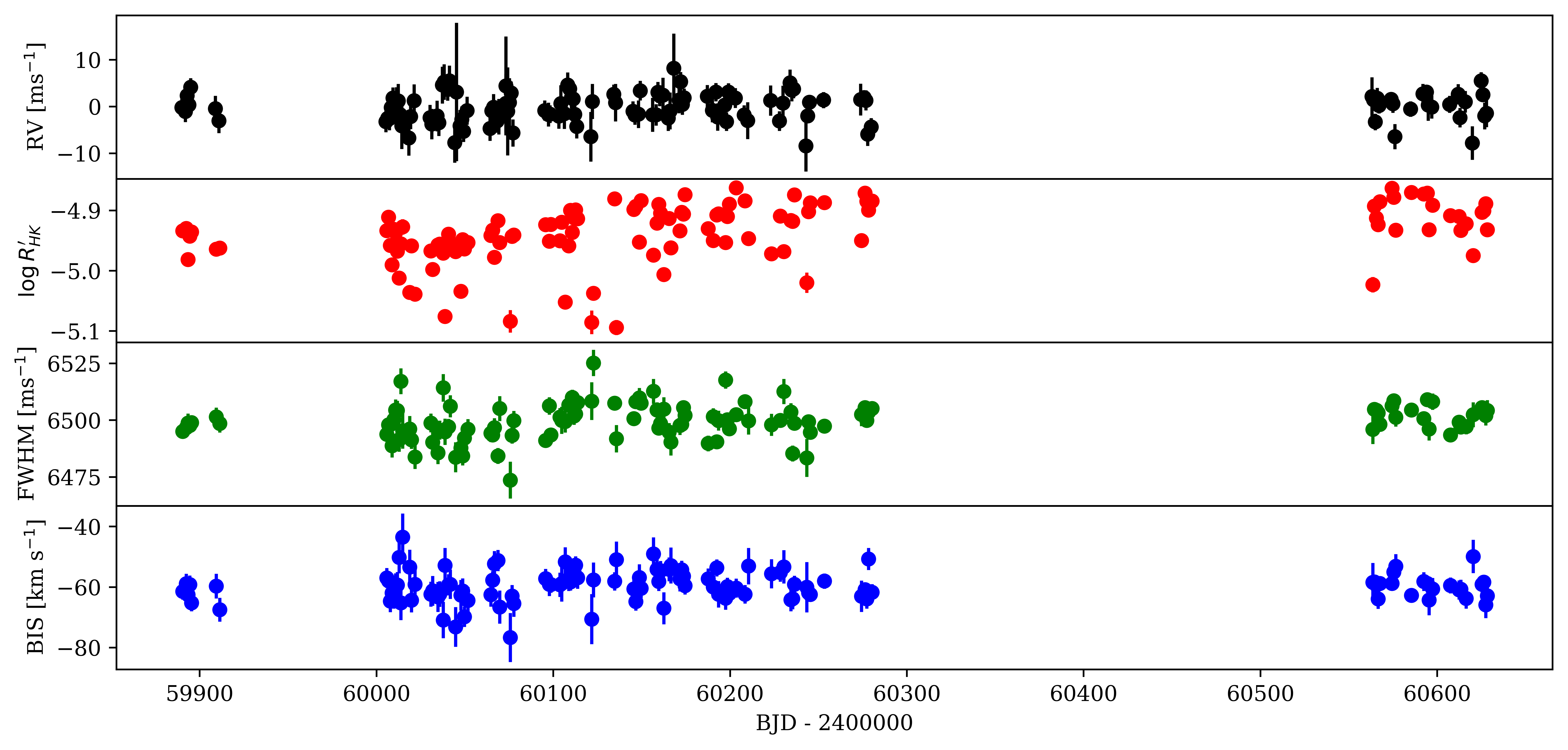}
    \caption{From top to bottom, time series of: MM-LSD RVs, \logrhk, CCF FWHM, and CCF BIS. 
    Five \logrhk observations have been omitted due to unphysical S-index measurements.
    The corresponding RV measurements and other activity indicators are retained.} 
    \label{fig:rv_data}
\end{figure*}

As both the anomalous calcium absorption and Fabry--P\'erot contamination cause spurious effects in isolated regions of the stellar spectra, we opt to also extract RVs using the multi-mask least-squares deconvolution method \citep[MM-LSD;][]{Lienhard22}. 
The benefits of MM-LSD over the CCF-based method of the DRS are twofold. 
Firstly, using the stellar parameters derived in Section \ref{sec:stellar_parameters}, we obtain a custom stellar mask from the Vienna Atomic Line Database \citep{VALD, VALD2a, Vald2b, VALD2c, Vald3}.
Secondly, the MM-LSD algorithm identifies anomalous spectral regions and masks them from all spectra in the time series. 
This means that, unlike in the DRS which calculates RVs on a spectrum-by-spectrum basis, a consistent list of good lines are used to derive RVs using MM-LSD.
To extract the optimal LSD profile, MM-LSD explores a grid of 32 hyperparameters\footnote{MM-LSD explores combinations of the following hyperparameters: the threshold to omit pixels due to telluric correction, the minimum line strength to consider, the maximum allowed model deviation, and the width of the LSD profile.} combinations to generate 32 individual RV time series. 
Whilst minimising the RMS is not necessarily the optimal metric for performance of MM-LSD when applied to planet-hosting stars, we opt to take the mean of the 16 time series with the lowest RMS \citep[see][]{Lienhard22}.
The final RV time series has an RMS of 3.29\,\si{\metre\per\second} with a median RV measurement uncertainty of 2.76\,\si{\metre\per\second}.
This is a 11 per cent reduction compared to the 3.66\,\si{\metre\per\second} RMS for the DRS-derived RVs.
We show the radial velocities, \logrhk, CCF FWHM, and BIS in Fig.~\ref{fig:rv_data}.
As MM-LSD is robust to spurious effects in isolated regions of spectra, we opt to retain the RVs whose $S$-index were rejected for being anomalous.

\begin{table*}
    \centering
        \caption{Summary of stellar parameters for TOI-5788.
    References [1] \citet{2021ApJS..254...39G}, [2] \citet{2018AJ....156..102S}, [3] \citet{2MASS}, [4] \citet{2021A&A...649A...1G}, [5] \citet{2000A&A...355L..27H}.}

    \begin{tabular}{cccc}
    \hline \hline
    Parameter & Description& Value & Reference \\
    \hline
    TOI & TESS Object of Interest & 5788 & [1]\\
    TIC & TESS Input Catalogue & 42883782 & [2]\\
    2MASS & $\cdots$ & J19094195+3145395 & [3]\\
    Gaia DR3 & $\cdots$ & 2042653052618584960 & [4]\\
    \hline
    $\alpha_\mathrm{J2000}$ & Right Ascension (RA) & 19h~09'~41.96"& [4]\\
    $\delta_\mathrm{J2000}$ & Declination (DEC) & +31$\si{\degree}$~45'~39.64"&[4]\\
    $\mu_\alpha$ & Proper motion (RA, mas~yr$^{-1}$) & 17.353&[4]\\
    $\mu_\delta$ & Proper motion (DEC, mas~yr$^{-1}$) & 119.563&[4]\\
    $\text{RV}_0$ & Absolute radial velocity (\si{\kilo\metre\per\second}) & $-59.36\pm0.46$ & [4]\\ 
    $\varpi$ & Parallax [$\si{\arcsecond}$]& $10.2688\pm$0.0123 & [4]\\
    $U$ & Galactocentric velocity (\si{\kilo\metre\per\second}) & $-75.03 \pm 0.21$& This work\\
    $V$ & Galactocentric velocity (\si{\kilo\metre\per\second}) & $-31.19 \pm 0.41$&This work\\
    $W$ & Galactocentric velocity (\si{\kilo\metre\per\second}) & $5 .69\pm 0.085$ & This work\\
    \hline
    $B$ & $B$ magnitude& $10.83\pm0.04$ & [5] \\
    $V$ & $V$ magnitude& $10.15\pm0.03$ & [5]\\
    $G$ & $G$ magnitude& $10.060\pm0.0028$ & [4]\\
    $J$ & $J$ magnitude & $8.950\pm0.026$ & [3]\\
    $H$ & $H$ magnitude & $8.657\pm0.033$ & [3]\\
    $K$ & $K$ magnitude & $8.577\pm0.017$ & [3]\\

    \hline
    $T_\mathrm{eff}$ & Effective temperature (K) & $5615\pm25$& This work\\
    $M_\star$ & Stellar mass ($\mathrm{M_\odot}$) & $0.87\pm 0.04$ & This work\\
    $R_\star$ & Stellar radius ($\mathrm{R_\odot}$) & $0.87\pm0.006$ & This work\\
    $\rho_\star$ & Stellar density ($\mathrm{\rho_\odot}$) & 1.34$^\pm0.09$ & This work\\
    Age& Stellar age (Gyr) & 5.72$^{+3.37}_{-2.65}$& This work\\
    $\left[\text{Fe/H}\right]$ & Metallicity (dex) &  $-0.32\pm0.04$ & This work\\
    $L_\star$& Luminosity ($\mathrm{L_\odot}$) & 0.712$\pm0.018$&This work\\
    $\log g_\mathrm{spec}$ & Surface gravity (cgs units) & 4.42$^\pm0.01$&This work\\
    $\log g_\mathrm{iso}$ & Surface gravity (cgs units) & 4.50$^{+0.025}_{-0.029}$&This work\\
    $v \sin i$& Projected rotational velocity (\si{\kilo\metre\per\second}) &  $\leq 2$ & This work\\
    $\xi$  & Microturbulence (\si{\kilo\metre\per\second}) & $ 0.93\pm 0.07$ & This work\\
    \logrhk & Chromospheric emission ratio& $-4.94 \pm 0.05$ & This work\\
\hline \hline
     \end{tabular}
    \label{tab:toi5788_parameters}
\end{table*}

\section{Stellar characterisation} \label{sec:stellar_characterisation}

\subsection{Stellar parameters} \label{sec:stellar_parameters}

TOI-5788 (TIC 42883782 in the \textit{TESS} input catalogue) is a relatively bright, high-proper-motion G dwarf star located at a distance of 97.4$\pm0.1$~pc \citep{gaiacollaborationGaiaMission2016, GaiaDR3}.
We derive galactic velocities of the TOI-5788 system using the method of \citet{JohnsonSoderblom} and data from the third Gaia data release \citep{gaiacollaborationGaiaMission2016, GaiaDR3}.
We report the galactocentric velocities in Table \ref{tab:toi5788_parameters}. 
Using the method of \citet{Reddy06}, we find that TOI-5788 has a 96.5 per cent probability of belonging to the Galactic thin disc.

Stellar atmospheric parameters for this star were derived from spectroscopic observations via three independent methods: ARES+MOOG, \textsc{CCFPams}, and SPC.
ARES+MOOG  \citep{2014dapb.book..297S} is a curve-of-growth method of obtaining stellar spectroscopic properties, using the equivalent widths (EWs) of iron spectral lines. 
The method has two main components. 
ARES measures the EWs from observed spectra, and then the radiative transfer code MOOG is used to calculate the individual abundances.
A stellar atmospheric model is adopted based on the atmospheric parameters.
By ensuring excitation and ionisation balance is achieved for all the analysed lines, we can calculate refined atmospheric parameters.
This analysis gave estimates for $T_\mathrm{eff}$, $\log g$, microturbulence $\xi$, and [Fe/H].
The surface gravity was corrected following \citet{2014A&A...572A..95M}.
A second method used is \textsc{CCFpams} \citep{2017MNRAS.469.3965M}.
This approach uses several CCFs to estimate $T_\mathrm{eff}$, $\log g$, [Fe/H] via an empirical calibration to literature values.
Finally, the Stellar Parameter Classification \citep[SPC;][]{2012Natur.486..375B, 2014Natur.509..593B} tool was used on individual HARPS-N spectra. 
This approach differs from the previous two in that SPC uses the entire spectrum between 5050$\,\si{\angstrom}$ and 5360$\,\si{\angstrom}$ to compare to a library of synthetic spectra. 
In addition to the  $T_\mathrm{eff}$, $\log g$, and [Fe/H] that ARES+MOOG and \textsc{CCFpams} provide, SPC also estimates the projected rotational velocity of the star ($v \sin i$).
The stellar parameters obtained by AREES+MOOG, \textsc{CCFpams}, and SPC are consistent within uncertainties.
As final parameters, we adopt the inverse-variance weighted average of the three methods.

Effective temperature and metallicity from these pipelines were then fed into the \textsc{isochrones} package \citep{2015ascl.soft03010M} and compared with stellar evolution models, in order to provide estimates of the mass, radius, and age of the star.
In particular, the stellar atmospheric parameters were compared to models from the Dartmouth Stellar Evolution Database \citep[DSED;][]{2008ApJS..178...89D} and the MESA Isochrones and Stellar Tracks \citep[MIST;][]{2016ApJS..222....8D}.
See \citet{Mortier2020} for a more in-depth explanation.
The adopted stellar parameters are listed in Table~\ref{tab:toi5788_parameters}.

\subsection{Individual chemical abundances}

Iron-poor stars are known to exhibit a wide range of chemical abundances, making them valuable test beds to study the link between stellar chemical abundances and planetary chemical composition \citep[][Turner et al. \textit{in prep}]{Adibekyan21}.
As TOI-5788 falls into this interesting region of parameter space, we extract individual abundances for a selection of atomic species. We again use ARES to extract the EWs of the different spectral lines and then MOOG together with ATLAS9 model atmospheres \citep{Kurucz1993} to calculate the individual chemical abundances \citep[see e.g. ][]{Adibekyan2012,Mortier2020}. 
We report the stellar abundances for a number of species in Table~\ref{tab:abundances}. 
We also quote a combined $[\alpha/\mathrm{H}]$ and $[\alpha/\mathrm{Fe}]$, using Mg, Si, and Ti as a proxy for alpha elements. This shows that TOI-5788 is not alpha-enhanced.
We note that the $[\mathrm{Fe}/\mathrm{H}]$ quoted in Table~\ref{tab:abundances} differs slightly from the value quoted in Table~\ref{tab:toi5788_parameters}, though the two values are consistent to within $1 \sigma$. 
This is because, of the methods outlined in Section~\ref{sec:stellar_parameters}, only ARES+MOOG is capable of calculating individual species abundances. 
In this Section, for consistency, we compare the $\alpha$-element abundances to the value of $[\mathrm{Fe}/\mathrm{H}]$ derived from ARES+MOOG rather than the weighted average adopted in general. 
For clarity, all further references to metallicity in this Paper are to the value presented in Table~\ref{tab:toi5788_parameters}.

\begin{table}
    \begin{center}
    \caption{Individual species abundances from ARES+MOOG corresponding to the best-fit stellar atmosphere model from that method.
    The [Fe/H] value quoted here differs slightly from the value quoted in Table~\ref{tab:toi5788_parameters} as we quote here the ARES+MOOG-derived value of [Fe/H] for consistency, whereas in Table~\ref{tab:toi5788_parameters} we quote the ensemble value of the different methods outlined in Section~\ref{sec:stellar_parameters}.}
    \begin{tabular}{cc}
    \hline\hline
          Species& Relative abundance  \\
          \hline
        $[\mathrm{Na}/\mathrm{H}]$		& $-0.177\pm0.15$	\\
        $[\mathrm{Mg}/\mathrm{H}]$	&	$-0.273	\pm0.13$	\\
        $[\mathrm{Al}/\mathrm{H}]$	&	$-0.213	\pm0.013$ \\
        $[\mathrm{Si}/\mathrm{H}]$	&	$-0.256\pm0.049$	\\
        $[\mathrm{Ca}/\mathrm{H}]$	&	$-0.229\pm0.043$	\\
        $[\mathrm{Sc}/\mathrm{H}]$	&	$-0.221\pm0.083$	\\
        $[\mathrm{ScII}/\mathrm{H}]$	&	$-0.271\pm0.051$	\\
        $[\mathrm{Ti}/\mathrm{H}]$	&	$-0.208\pm0.042$	\\
        $[\mathrm{TiII}/\mathrm{H}]$	&	$-0.261\pm0.081$ \\
        $[\mathrm{Mn}/\mathrm{H}]$	&	$-0.326\pm0.054$	\\
        $[\mathrm{Cr}/\mathrm{H}]$	&	$-0.264\pm0.032$	\\
        $[\mathrm{CrII}/\mathrm{H}]$	&	$-0.195	\pm0.155$	\\
        $[\mathrm{V}/\mathrm{H}]$	&	$-0.217\pm0.055$	\\
        $[\mathrm{Co}/\mathrm{H}]$	&	$-0.276\pm0.043$	\\
        $[\mathrm{Ni}/\mathrm{H}]$	&	$-0.293\pm0.044$	\\
        $[\mathrm{C}/\mathrm{H}]$	&	$-0.266\pm0.206$	\\
        $[\mathrm{S}/\mathrm{H}]$	&	$-0.533\pm0.179$	\\
        \hline 
        $[\mathrm{Fe}/\mathrm{H}]$ & $-0.290\pm 0.05$ \\
        $[\alpha/\mathrm{H}]$ &  $-0.231 \pm 0.03$ \\
        $[\alpha/\mathrm{Fe}]$ & $0.059 \pm 0.08$\\
        \hline\hline
    \end{tabular}
    \label{tab:abundances}
    \end{center}
\end{table}

\subsection{Stellar activity} \label{sec:stellar_activity}

In order to rule out planetary false positives, it is important to accurately characterise the stellar activity signal, and measure the stellar rotation period.
We first estimate a maximum rotation period from the $v\sin i$ measurement.
We calculate the maximum rotation period as 
\begin{equation}
    P_\mathrm{rot, max} = \frac{2 \pi R_\star}{v\sin i}.
\end{equation}
Taking the upper limit $v\sin i$ measurement of 2\,\si{\kilo\metre\per\second}, we obtain an upper-bound rotation period of $P_\mathrm{rot, max} = 22.0\,\si{\day}$.
We note that, since we are not sensitive to measurements of $v\sin i$ below $2\si{\kilo\metre\per\second}$, that this value of $P_\mathrm{rot, max}$ should be interpreted with caution.

We also follow the method of \citet{1984ApJ...279..763N} to estimate the rotation period from the measured average chromospheric emission.
Our average \logrhk of $-4.94 \pm 0.05$ yields a period estimate of 27.2$\,\si{\day}$. 

\begin{figure*}
    \centering
    \includegraphics[width=1\linewidth]{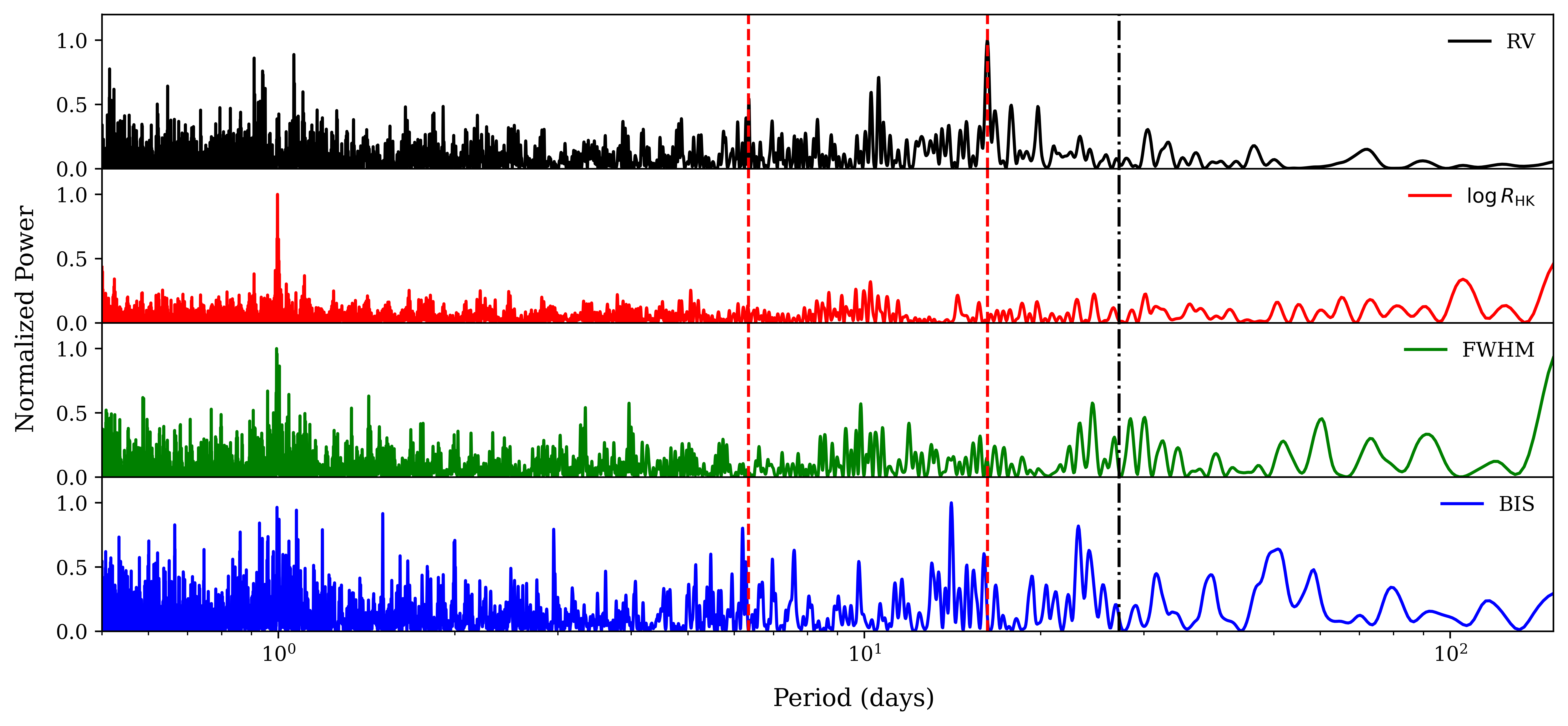}
    \caption{Normalised BGLS periodograms for the radial velocities, S-index, CCF full width at half maximum, and CCF bisector inverse slope.
    The periods of the planets are shown with red dashed vertical lines, and the predicted rotation period from \citet{1984ApJ...279..763N} is shown with a black dot-dashed vertical line.}
    \label{fig:bgls_periodograms}
\end{figure*}

Finally, to empirically estimate the rotation period as measured in the data, we compute Bayesian Generalised Lomb-Scargle \citep[BGLS][]{MortierBGLS} periodograms for the radial velocity data and activity indicators shown in Fig.~\ref{fig:rv_data}.
We show the periodograms in Fig.~\ref{fig:bgls_periodograms}.
The dashed vertical line shows the 27.2$\,\si{\day}$ period estimate.
We see no significant peak in either the RV or activity indicator periodograms. 
Owing to the low level of activity in TOI-5788, we are unable to robustly detect a rotation period through activity indicators.

\section{Transit analysis} \label{sec:transit_analysis}

The \textit{TESS} and \textit{CHEOPS} photometric data (Sections \ref{sec:tess_data} and \ref{sec:cheops_data}) were modelled jointly with a two-planet model using \textsc{Allesfitter} \citep{allesfitter-code, allesfitter-paper}. 
The limb darkening of TOI-5788 is modelled using the model of \citet{2013MNRAS.435.2152K}:
\begin{equation} \label{eq:limb_dark}
    \frac{I(\mu)}{I_0} = 1 - q_1 (1 - \mu) - q_2 (1 - \mu)^2,
\end{equation}
where $q_1$ and $q_2$ are the limb-darkening parameters, $I_0$ is the intensity at disc centre, and $\mu$ is the cosine of the angle between the normal to the stellar surface and the line of sight.
Limb-darkening parameters are fit separately for the \textit{TESS} 600-second cadence data, TESS 120-second cadence data, \textit{TESS} 20-second cadence data, and \textit{CHEOPS} data.
Priors for the limb-darkening coefficients were centred on the values derived in \citet{2000A&A...363.1081C}.

Priors for the planet orbital periods and times of central transit were derived from the automated pipeline from \textit{TESS}.
Uninformative priors were applied for the star-to-planet radius ratio and orbital inclination, requiring a transit depth of less than 0.1 per cent and a geometry that allows the planet to transit.
We parameterise $e$ and $\omega$ following \citet{2011ApJ...726L..19A,Eastman13}, i.e. as $\sqrt{e} \sin\omega$ and $\sqrt{e}\cos\omega$. 
Priors were chosen to match the eccentricity prior of \citet[][i.e., a zero-mean half-Gaussian with $\sigma  = 0.083$]{VanEylen19}.
We explored the parameter space with the \textsc{dynesty} dynamic nested sampler \citep{2020MNRAS.493.3132S} with 500 live points. 
We show the best-fit parameters in Table~\ref{tab:toi_transit_params}.

\begin{table}
\begin{center}
\caption{Best-fit orbital parameters from the transit analysis in Section~\ref{sec:transit_analysis}. The choice of priors is described in the text.}
\begin{tabular}{ccc}
\hline \hline 
Parameter& Prior& Value\\
\hline
$P_\mathrm{b}$ [d]& $\mathcal{U}[6.3, 6.4]$&  $6.340804 \pm 0.000018$\\
$R_\mathrm{b} / R_\star$& $\mathcal{U}[0, 0.3162]$& $0.01610 \pm 0.00076 $\\
$T_{0, \mathrm{b}}$ [$\mathrm{MJD} - 59600$]& $\mathcal{U}[11.232008, 11.250608]$& $11.2404 \pm 0.0019$ \\
$e_\mathrm{b}$ & $\mathcal{N^+}[0, 0.083] $ & $0.042^{+0.038}_{-0.030}$ \\[0.75ex]
$\cos{i_\mathrm{b}}$&  $\mathcal{U}[0, 0.0523]$& $0.0359 ^{+0.0034}_{-0.0046}$\\[0.75ex]
$(R_\mathrm{b} + R_\star)/a$ & $\mathcal{U}[0.0589, 0.0677]$ & $0.06391 ^\pm 0.001$\\
\hline
$P_\mathrm{c}$ [d]& $\mathcal{U}[16.2, 16.3]$&  $16.213358^{+0.000052}_{-0.000041}$\\[0.75ex]
$R_\mathrm{c} / R_\star$& $\mathcal{U}[0, 0.3162]$& $0.02394 \pm 0.00037$\\
$T_{0, \mathrm{c}}$ [$\mathrm{MJD} - 59600$]& $\mathcal{U}[8.082185, 8.095085]$& $8.0906 \pm 0.0012$ \\
$e_\mathrm{c}$ & $\mathcal{N^+}[0, 0.083] $ & $0.047^\pm0.033$ \\
$\cos{i_\mathrm{c}}$&  $\mathcal{U}[0, 0.0232]$& $0.007 \pm 0.004$\\
$(R_\mathrm{c} + R_\star)/a$ & $\mathcal{U}[0.03151, 0.03616]$ & $0.03454\pm 0.00057$\\
\hline

$q_{1\mathrm{, TESS600}}$&$\mathcal{U}[0.3322, 0.3522]$& $0.3415\pm0.006$\\
$q_{2\mathrm{, TESS600}}$&$\mathcal{U}[0.3147, 0.3347]$& $0.3259\pm0.006$\\
$q_{1\mathrm{, TESS120}}$&$\mathcal{U}[0.3322, 0.3522]$& $0.3418\pm0.006$\\
$q_{2\mathrm{, TESS120}}$&$\mathcal{U}[0.3147, 0.3347]$& $0.3238\pm0.006$\\
$q_{1\mathrm{, TESS020}}$&$\mathcal{U}[0.3322, 0.3522]$& $0.3422\pm0.006$\\
$q_{2\mathrm{, TESS020}}$&$\mathcal{U}[0.3147, 0.3347]$& $0.3293^{+0.0037}_{-0.0053}$\\
$q_{1\mathrm{, CHEOPS}}$&$\mathcal{U}[0.4768, 0.4968]$& $0.4862\pm0.006$\\[0.75ex]
$q_{2\mathrm{, CHEOPS}}$&$\mathcal{U}[0.3741, 0.3941]$& $0.3863^{+0.0051}_{-0.0066}$\\
\hline\hline
\end{tabular}
\end{center}
\label{tab:toi_transit_params}
\end{table}%

\section{Radial-velocity analysis} \label{sec:rv_analysis}

In this Section, we detail our analysis of the radial-velocity data described in Section~\ref{sec:rv_data}.
We analysed the radial-velocity data with the \textsc{pyorbit} python package \citep{Malavolta16, Malavolta18}.
Since the transit fit of Section~\ref{sec:transit_analysis} indicates low eccentricities, we consider both a circular, and full Keplerian orbit for each planet to give a total of four models.

Because the periods of both planets are well constrained by the transit analysis, we enforce strict Gaussian priors on the periods and times of central transit using the values in Table~\ref{tab:toi_transit_params}.
For the Keplerian fits, we adopt eccentricity priors from \citet{VanEylen19}; a zero-mean half-Gaussian with $\sigma = 0.083$.
We provide uninformative priors for RV semi-amplitude and, where applicable, argument of periastron.
As we have well-constrained mid-transit times from the transit analysis of Section~\ref{sec:transit_analysis}, we opt to use the parameterisation of \citet{Eastman13}. 
In this formalism, we sample the Keplerian orbital parameters in the following parameter space: $\log P$, $\log K$, $\sqrt{e}\sin{\omega}$, $\sqrt{e}\cos{\omega}$, $T_c$.
This choice avoids a boundary condition at $e=0$.
We also fit the RV offset and jitter, with uninformative priors.

We sample the model parameter space using the \textsc{dynesty} nested-sampler algorithm as the calculation of the log-evidence allows for direct model comparison.
We find that no model is particularly preferred. 
Therefore for generality, we adopt a two-Keplerian model with eccentricity as a free parameter.

\begin{figure}
    \centering
    \includegraphics[width=1\linewidth]{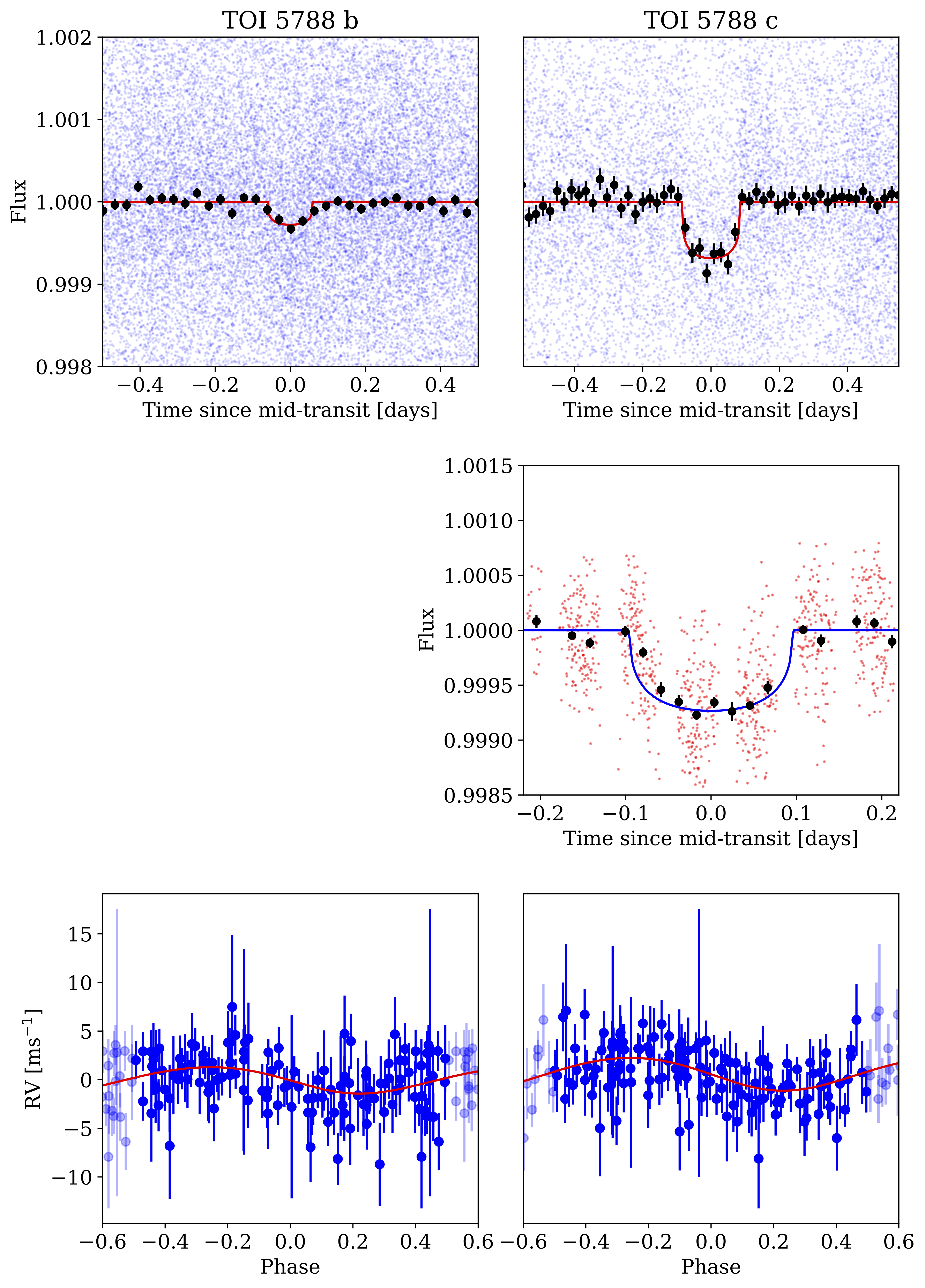}
    \caption{(Top) Phase-folded \textit{TESS} transits light curves of TOI-5788~b and TOI-5788~c. 
    The black points show the binned transit to emphasise the transit depths of roughly 0.02 per cent and 0.06 per cent, respectively.
    The standard errors for the phase-binned observations are shown but are not visible on the scale of the plot.
    (Middle) Phase-folded \textit{CHEOPS} transit light curve of TOI-5788~c. As above, black points show the phase-binned light curve with standard errors. TOI-5788~b was not observed by \textit{CHEOPS}.
    (Bottom) The phase folded radial velocity curves. 
The best-fit transit and radial-velocity models are shown as solid lines.}
    \label{fig:phase_fold}
\end{figure}

\subsection{Gaussian processes}

In addition to the adopted model of the previous Section, we also consider a two-Keplerian model with a noise model to account for stellar activity. 
We model the activity as a quasi-periodic Gaussian process \citep[GP;][]{haywoodPlanetsStellarActivity2014, Rajpaul15}.
This GP models the covariance between two observations as
\begin{equation}
    K(t_1, t_2) = A^2 \exp\left[ \Gamma \sin^2\left(\frac{\pi (t_1 - t_2)}{P_\mathrm{rot}} \right) - \frac{(t_1 - t_2)^2}{\tau^2}\right],
\end{equation}
where the hyperparameters $A$, $\Gamma$, $P_\mathrm{rot}$, and $\tau$ encode the GP amplitude, harmonic complexity, rotation period, and decay timescale, respectively.

Since we do not find a preferred rotation period in Section~\ref{sec:stellar_activity}, we provide broad, uninformative priors for all parameters.
We again sample the parameter space using \textsc{dynesty}. 
We find that the model with two Keplerians and a GP is moderately disfavoured compared to the activity-free model ($\Delta \log Z = - 2.647)$. 
We also find that the GP amplitude is consistent with zero ($ 0.57_{ -0.39}^{+0.46}\,\si{\metre\per\second}$).
Finally, we obtain the same masses for the two planets in the GP model as in the activity-free model, with no improvement in precision. 
For these reasons, we opt to use the two-Keplerian, activity-free model as the best-fit model for the masses.
We show the best-fit orbital parameters in Table~\ref{tab:rv_parameters}.
This is consistent with the low values of \logrhk and the fact we find no clear periodicities in any activity indicator in Section~\ref{sec:stellar_characterisation}.

\begin{table}

\begin{center}
\caption{Best-fit orbital parameters from the radial-velocity analysis in Section~\ref{sec:rv_analysis}. The choice of priors is described in the text.}
\begin{tabular}{ccc}
\hline \hline 
Parameter& Prior& Value\\
\hline
$P_\mathrm{b}$ [d]& $\mathcal{N}[6.340804, 0.000018]$&  $6.340804 \pm 0.000018$\\
$K_\mathrm{b}$ [$\si{\metre\per\second}$]& $\mathcal{U}[0.01, 300.0]$& $1.42\pm0.36 $\\
$T_{0, \mathrm{b}}$ [MJD-59600]& $\mathcal{N}[11.2426, 0.002]$& $ 11.2425\pm 0.002$\\
$e_\mathrm{b}$ & $\mathcal{N^+}[0, 0.083] $ & $0.061^{+0.066}_{-0.042}$ \\[0.75ex]
$\omega_\mathrm{b}$ [deg]&  $\mathcal{U}[0, 360]$& $96 ^{+75}_{-116}$\\[0.75ex]
\hline
$P_\mathrm{c}$ [d]& $\mathcal{N}[16.213358, 0.000045]$&  $16.213357 \pm 0.000045$\\
$K_\mathrm{c}$ [$\si{\metre\per\second}$]& $\mathcal{U}[0.01, 300.0]$& $ 1.78_{-0.34}^{+0.33}$\\
$T_{0, \mathrm{c}}$ [MJD-59600]& $\mathcal{N}[8.0906, 0.0013]$& $ 8.0906\pm 0.0013$\\
$e_\mathrm{c}$ & $\mathcal{N^+}[0, 0.083] $ & $ 0.048_{-0.033}^{+0.051}$\\[0.75ex]
$\omega_\mathrm{c}$ [deg]&  $\mathcal{U}[0, 360]$& $148_{-116}^{+127}$\\[0.75ex]
\hline
$\sigma$ [$\si{\metre\per\second}$] & $\mathcal{U}[0.01, 1500]$ & $0.42_{-0.28}^{+0.39}$ \\[0.75ex]
$\mathrm{RV_0}$ [$\si{\metre\per\second}$] & $\mathcal{U}[-69517, -49498]$ & $-59508.58\pm0.24$\\
\hline\hline
\label{tab:rv_parameters}
\end{tabular}
\end{center}

\end{table}%

\subsection{Stability and additional planets in the system}

\begin{figure*}
    \centering
	\includegraphics[width=1\textwidth]{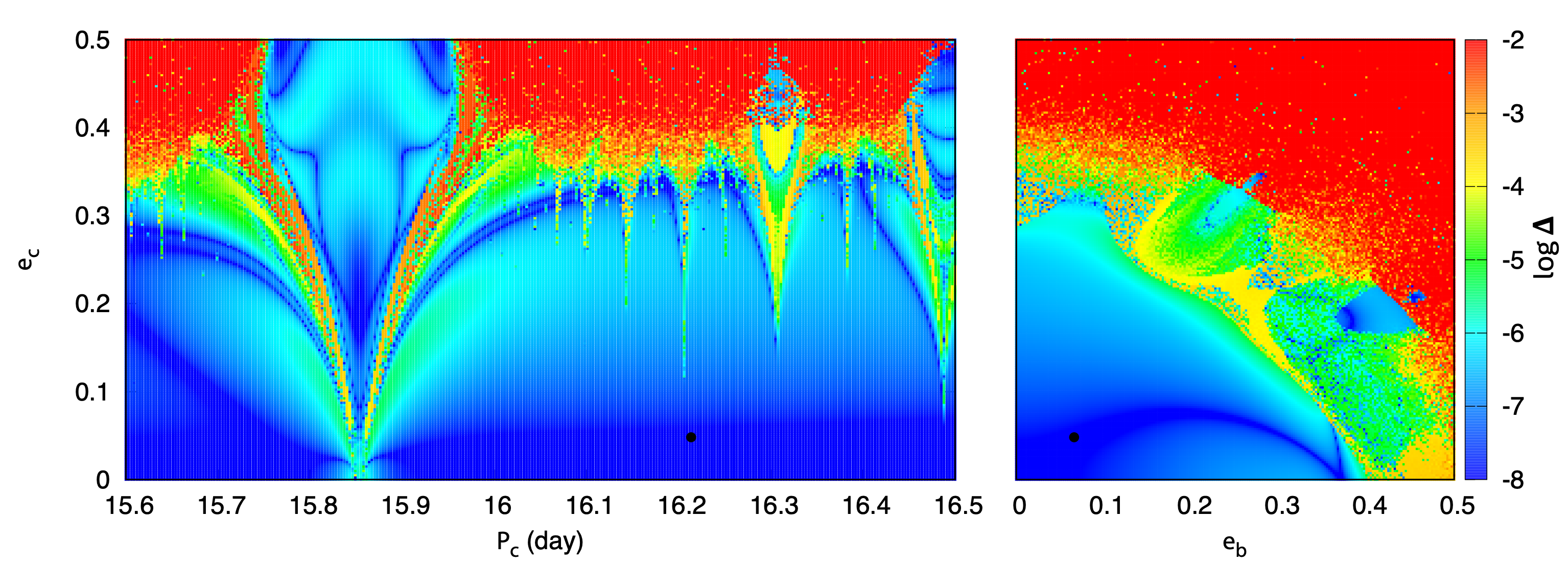}
    \caption{Stability analysis of the TOI-5788 planetary system. For fixed initial conditions (Table~\ref{tab:toi_transit_params}), the parameter space of the system is explored by varying the orbital period and the eccentricity of planet c (left panel) and the eccentricities of both planets (right panel). The step size is $0.0025$ in the eccentricities, $0.001$\,d in the orbital period of planet b, and $0.003$\,d in the orbital period of planet c. 
    For each initial condition, the system is integrated over 5000\,yr and, from a frequency analysis of the mean longitude of the outer planet, a stability indicator is calculated.
    The colours in the plot represent the chaotic diffusion, measured by the variation in the frequency (see text).
    Red points correspond to highly unstable orbits, whilst blue points correspond to orbits which are likely to be stable on Gyr timescales. The black dots show the values of the best fit solution (Table~\ref{tab:toi_transit_params}).}
    \label{figSA}
\end{figure*}

The TOI-5788 system is composed of two close-in planets near a 5:2 mean motion resonance ($P_\mathrm{c} / P_\mathrm{b} \approx 2.56$, Table~\ref{tab:adopted_parameters}).
In order to get a clear view on the system dynamics, we performed a stability analysis in a similar way as for other planetary systems \citep[e.g.,][]{Correia_etal_2005, Correia_etal_2010}.
The system is integrated on a regular 2D mesh of initial conditions in the vicinity of the best fit (Table~\ref{tab:toi_transit_params}).
We used the symplectic integrator SABAC4 \citep{Laskar_Robutel_2001}, with a step size of $5 \times 10^{-4} $\,yr and general relativity corrections.
Each initial condition is integrated for 5000~yr, and a stability indicator, $\Delta = |1-n'/n|$, is computed. 
Here, $n$ and $n'$ are the main frequencies over two consecutive time intervals of 2500~yr, calculated as in \citet{Laskar_1990, Laskar_1993PD}. 
The results are shown in Figure \ref{figSA}: orange and red represent strongly chaotic unstable trajectories; yellow indicates the transition between stable and unstable regimes; green corresponds to moderately chaotic trajectories (stable on Gyr timescales); cyan and blue give extremely stable quasi-periodic orbits.

We first explore the stability of the system by varying the orbital period and the eccentricity of the outer planet (Fig.~\ref{figSA}, left) and both eccentricities (Fig.~\ref{figSA}, right).
We observe that the best fit solution from Table~\ref{tab:toi_transit_params} is completely stable (black dots in Fig.~\ref{figSA}), even if we increase the eccentricities up to 0.3. 
In addition, we verify that the system is outside the 5:2 mean motion resonance, which corresponds to the large stable V-shape structure in the left figure.
We therefore conclude that the system parameters presented in Table~\ref{tab:toi_transit_params} correspond to a highly dynamically stable configuration.

\begin{figure*}
    \centering
	\includegraphics[width=\textwidth]{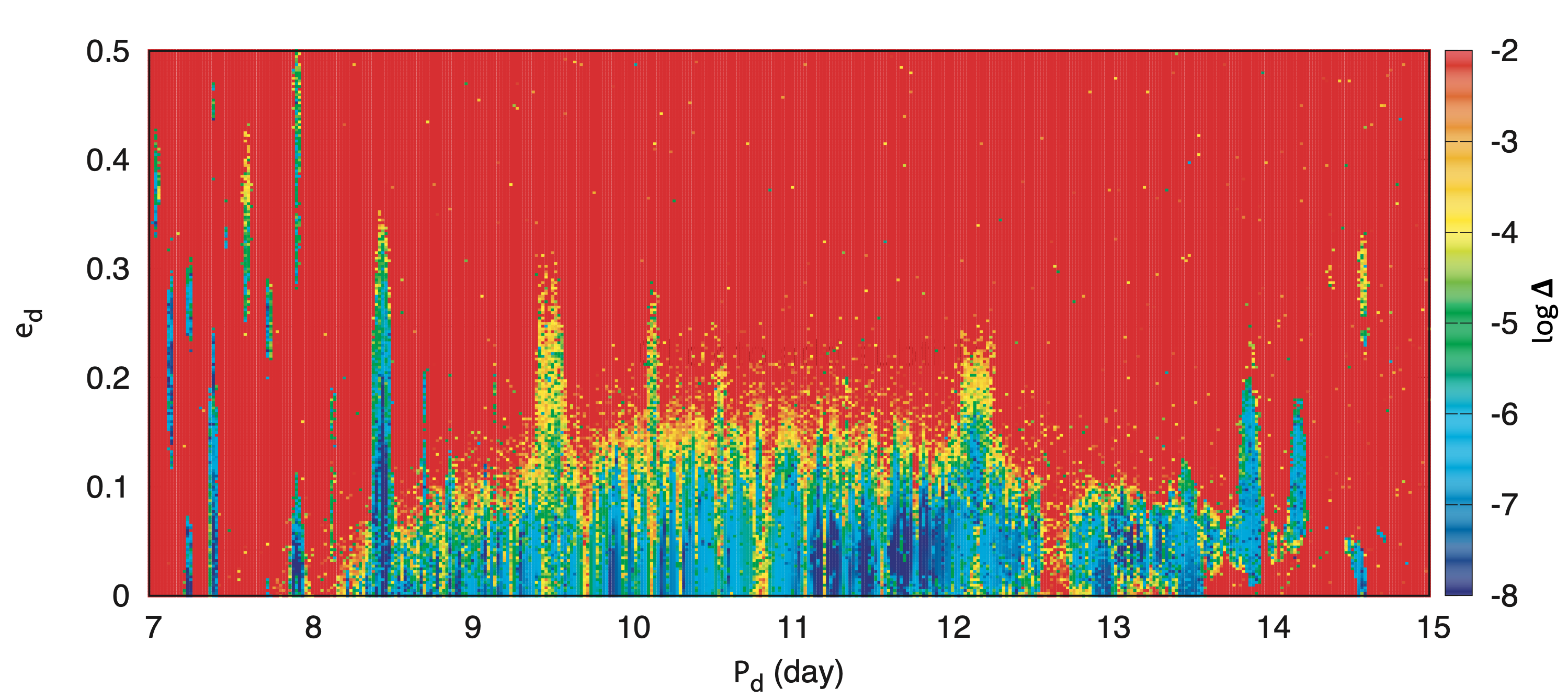}
    \caption{Stability analysis of a potential third planet, d, in the TOI-5788 system assuming $K = 0.5\,\mathrm{m\,s^{-1}}$ and coplanar orbits. For fixed initial conditions (Table~\ref{tab:toi_transit_params}), the parameter space of the system is explored by varying the orbital period $P_\mathrm{d}$ and the eccentricity $e_\mathrm{d}$ of the tentative planet d. The step size is $0.02$\,d in orbital period and $0.0025$ in eccentricity. The colour codes are the same as in Fig.~\ref{figSA}.}
    \label{figSB}
\end{figure*}

We then investigate the stability of a potential planet d between the orbits of planets b and c with a semi-amplitude of $K = 0.5$~m/s, which roughly corresponds to the HARPS-N current RV detection limit (Fig.~\ref{figSB}). 
We observe that the presence of such a planet is indeed possible between roughly 8 and 14 days, so we cannot rule out the existence of additional small mass planets in that region.
We ran a three-Keplerian model to search the 8--14 day region, but this was disfavoured compared to the two-Keplerian model we present in Table~\ref{tab:rv_parameters}.
Additionally, similar to the model that included a GP, the amplitude of the additional signal tended to the lowest allowed values ($M_\mathrm{d} = 0.37_{-0.30}^{+1.4}\,\mathrm{M_\oplus}$).
We therefore report no evidence of a third planet lying between planets b and c.

\subsection{Adopted parameters}
Having considered a number of models to describe the data, we adopt a two-Keplerian solution, with stellar activity being absorbed into the white-noise jitter term.
We report the adopted planetary parameters in Table~\ref{tab:adopted_parameters}.
Planetary parameters were derived from the parameters in Tables~\ref{tab:toi5788_parameters}, \ref{tab:toi_transit_params}, and \ref{tab:rv_parameters}. 
As the priors for radial-velocity analysis were informed by the transit analysis, we take the values of $P$ and $T_0$ from the radial-velocity analysis.
We also adopt the values of eccentricity from the radial-velocity analysis as, in the absence of secondary transits, radial-velocity signals allow for more the precise measurement of eccentricities.
The best-fit transit and RV models are shown in Fig~\ref{fig:phase_fold}.

\begin{table}
    \centering
    \caption{Adopted planetary parameters for the TOI-5788 system.}
    \begin{tabular}{ccc}
         \hline \hline 
         Parameter & TOI-5788~b & TOI-5788~c\\
         \hline
         $P$ [d] &$6.340758 \pm 0.000030$&$16.213362 \pm 0.000026$\\
         $T_0$ [MJD] &$ 59611.2425\pm 0.002$&$ 59608.091365\pm 0.0008$\\
         $M$ [$M_\oplus$]&$3.72\pm0.94$&$6.4\pm1.2$\\
         $R$ [$R_\oplus$]&$1.528\pm0.075$&$2.272\pm0.039 $\\
         $\rho$ [$\rho_\oplus$]&$1.04\pm0.28$&$0.55\pm0.10$\\
         $a$ [AU]&$0.0640\pm0.0011$&$0.1197\pm0.0020$\\[0.75ex]
         $i$ [deg] & $87.94^{+0.26}_{-0.20}$& $89.60\pm0.25$\\[0.75ex]
         $e$ & $0.061^{+0.066}_{-0.042}$& $ 0.046_{-0.032}^{+0.051}$\\[0.75ex]
\hline \hline
    \end{tabular}
    \label{tab:adopted_parameters}
\end{table}

\section{Planetary characterisation}\label{sec:characterisation}

\begin{figure}
    \centering
    \includegraphics[width=1\linewidth]{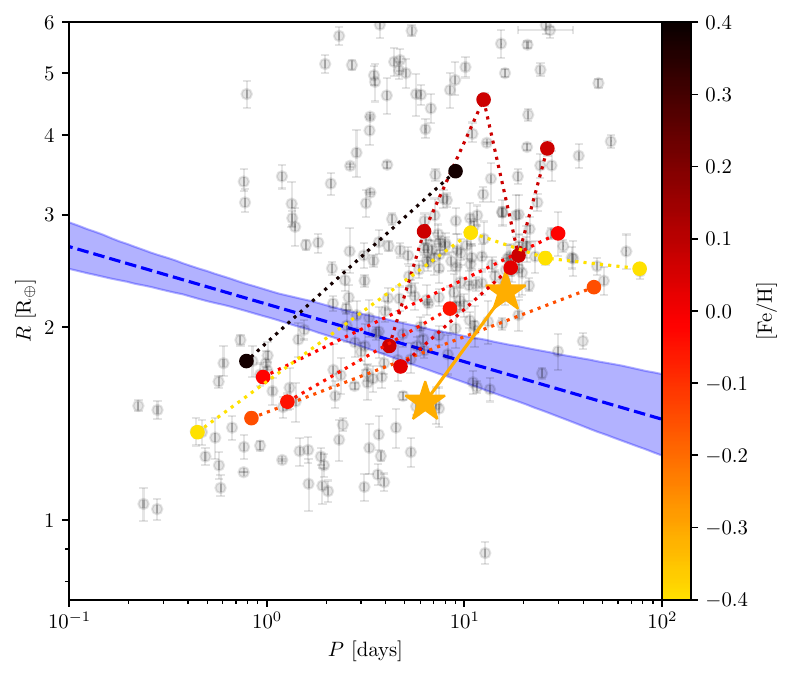}
    \caption{The period--radius relationship of known exoplanets.
    Grey points show planets with 10-$\sigma$ measurement on radius and radial-velocity masses to at least a 3-$\sigma$ detection.
    The radius valley of \citet{2023MNRAS.519.4056H} is shown in blue. 
    Points joined by dotted lines show planets in multi-planet systems, with at least one planet either side of the radius valley, and are coloured by host-star metallicity.
    The TOI-5788 planets are shown by the stars. 
    Of these systems, the TOI-5788 planets straddle the radius valley the closest, and are only the second most metal-poor. 
    This makes the TOI-5788 system a vital system for exoplanet formation studies.
    }
    \label{fig:radius_valley}
\end{figure}

The TOI-5788 system occupies an interesting region of parameter space, with planets spanning the radius valley. 
To date 87 such systems have been discovered, of which only 32 have the planets' physical properties well-constrained\footnote{Data from \url{www.exoplanet.eu} - 2025-07-16.}\footnote{Here we consider a planet to have well constrained properties if its radius is known to 10 per cent, and it has a radial velocity mass to at least a $3-\sigma$ detection}.  
The majority of these planets orbit cool stars or stars with supersolar metallicity. 
In Fig.~\ref{fig:radius_valley}, we plot the period-radius distribution of all known exoplanets smaller than six Earth radii with a 10-$\sigma$ measurement on radius and radial-velocity masses to at least a 3-$\sigma$ detection, as well as the period-dependent radius gap described by \citet{2023MNRAS.519.4056H}.
We also overlay planetary systems wherein Sun-like stars ($5000\,\si{\kelvin} \leq T_\mathrm{eff} \leq 6000\,\si{\kelvin}$) hosting planets that span the radius gap. 
Of these systems, TOI-5788 is one of the only significantly metal-poor host stars, and its planets straddle the radius gap more closely than the other systems.

Metal-poor stars are known to display a wider range of alpha-element enhancement than metal-rich stars \citep{Bensby2003, Adibekyan2012, 2025PASA...42...51B}.  
This makes metal-poor exoplanet hosts vital to study the link between stellar and planetary compositions.
The unique position of TOI-5788 as the only significantly metal-poor solar-type star with well-characterised planets which closely straddle the radius valley therefore makes the system a valuable test bed for super-Earth/sub-Neptune formation theories. 
In this section, we further characterise the physical properties of TOI-5788~b and TOI-5788~c.

\subsection{Instellation and equilibrium temperature} \label{sec:instellation}

Planetary instellation is given by
\begin{equation}
    F_\mathrm{p} = \frac{\sigma R_\star^2 T_\mathrm{eff}^4}{a^2} = \left( \frac{T_\mathrm{eff}}{5777 \si{\kelvin}}\right)^4 \left( \frac {R_\star}{\mathrm{R_\odot}}\right)^2 \left( \frac{a}{1\mathrm{AU}} \right)^{-2} \mathrm{F_\oplus},
\end{equation}
where $\sigma$ is the Stefan-Boltzmann constant, $a$ is the semi-major axis of the planet, and $\mathrm{F_\oplus}$ is the insolation of the Earth.

We find that the inner and outer planets have incident stellar fluxes of $F_\mathrm{b} = 175.1\pm6.7\,\mathrm{F_\oplus}$, and $F_\mathrm{c} = 49.8\pm2.0\,\mathrm{F_\oplus}$, respectively. 
We can also calculate their equilibrium temperatures as
\begin{equation}
    T_\mathrm{eq} = T_\mathrm{eff} \sqrt{ \frac{R_\star}{2a}} \left( 1 - A_\mathrm{B}\right) ^ {1/4},
\end{equation}
where the Bond albedo $A_\mathrm{B}$ quantifies the fraction of incident flux reflected by the planet.
For $A_\mathrm{B} = 0.3$, comparable to the majority of Solar System planets, this gives $T_{\mathrm{eq,b}} = 910\,\si{\kelvin}$ and $T_{\mathrm{eq, c}} = 667\,\si{\kelvin}$.

\subsection{Planetary composition}
\label{sec:comp}

\begin{figure}
    \centering
    \includegraphics[width=1\linewidth]{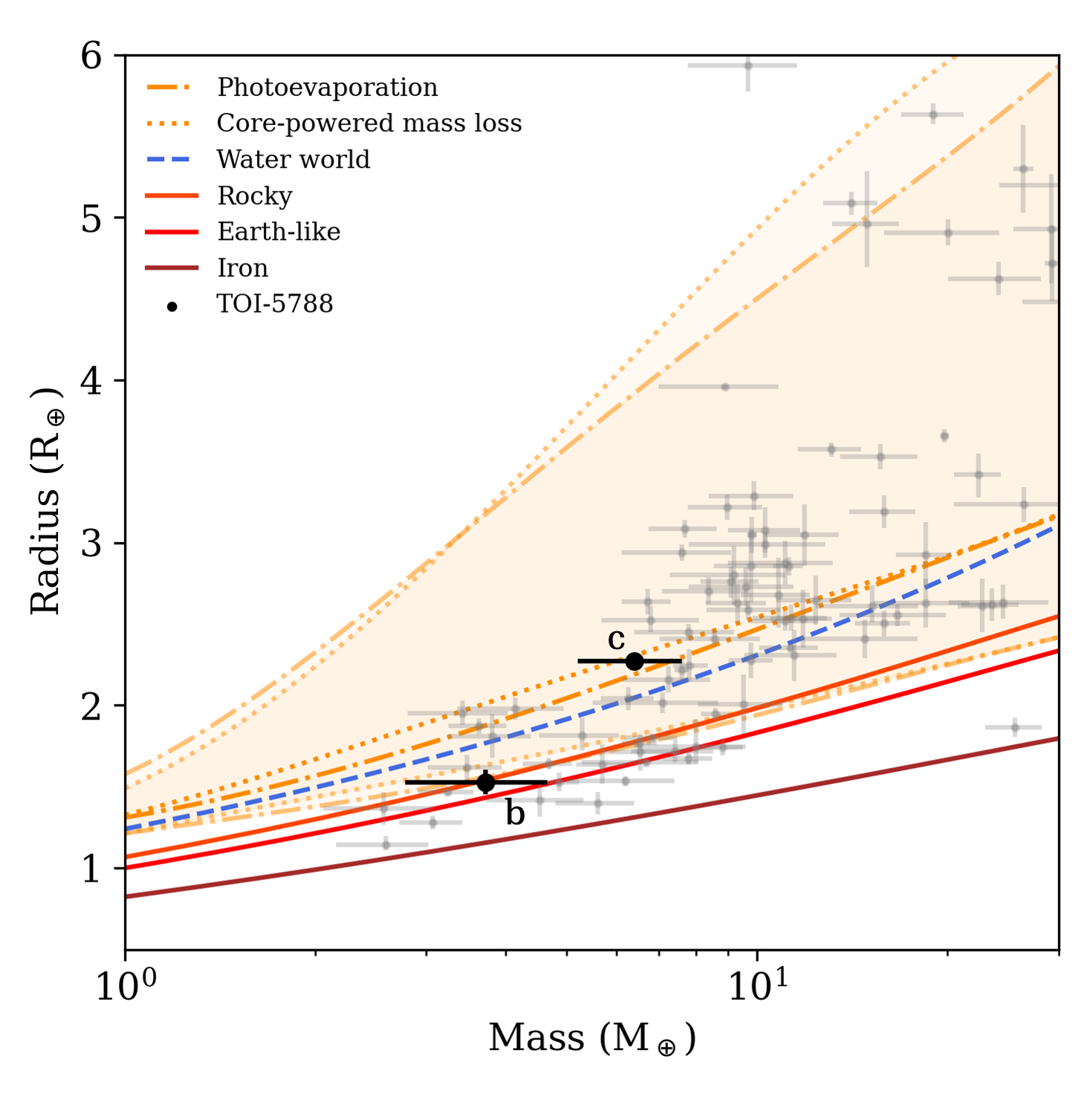}
    \caption{The mass--radius relationship for exoplanets with $R \leq 6\,\mathrm{R_\oplus}$.
    A number of theoretical mass--radius relationships are overlaid allowing for estimates of planet bulk composition.
    The dotted and dot-dashed orange lines show the core-powered mass-loss and photoevaporation models, respectively. 
    The central relations show the boil-off initial conditions, and the larger shaded areas show agnostic initial conditions \citep[see][for more details]{Rogers23}.
    The blue dashed line is the relation for a water-world with a 1:1 silicate-to-ice ratio \citep{2019PNAS..116.9723Z, 2022Sci...377.1211L}.
    The solid lines show the models for a rocky world (pure MgSiO$_3$), an Earth-like composition (32.5 per cent iron to 67.5 per cent MgSiO$_3$), and a pure-iron planet \citep{2019PNAS..116.9723Z}.
    Overlaid in grey are known planets with radius better than 10-$\sigma$ and RV mass measurements better than 3-$\sigma$.}
    \label{fig:mass_radius}
\end{figure}

\begin{figure*}
    \centering
    \includegraphics[width=1\linewidth]{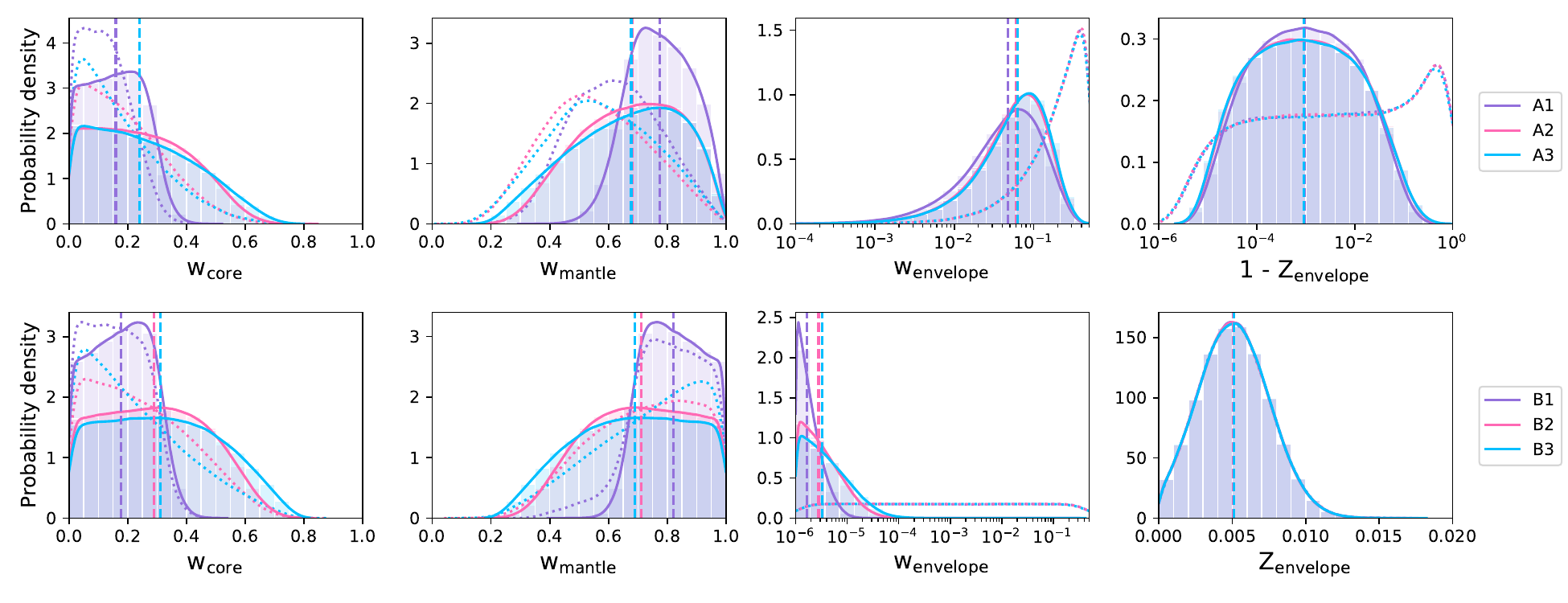}
    \caption{    
    Interior composition models of TOI-5788~b.
    Posterior (solid) and prior (dotted) distributions of the core, mantle, and envelope mass fractions along with the fraction of water in the envelope by mass ($Z_\mathrm{envelope}$) are shown for the six formation scenarios described in the text.
    Upper panels correspond to water-rich formation conditions and lower panels correspond to the water-poor case.
    Purple, pink, and blue distributions correspond to the stellar-like, iron-enriched, and free priors for silicate abundances, respectively. 
    }
    \label{fig:toi_5788b_interior}
\end{figure*}

\begin{figure*}
    \centering
    \includegraphics[width=1\linewidth]{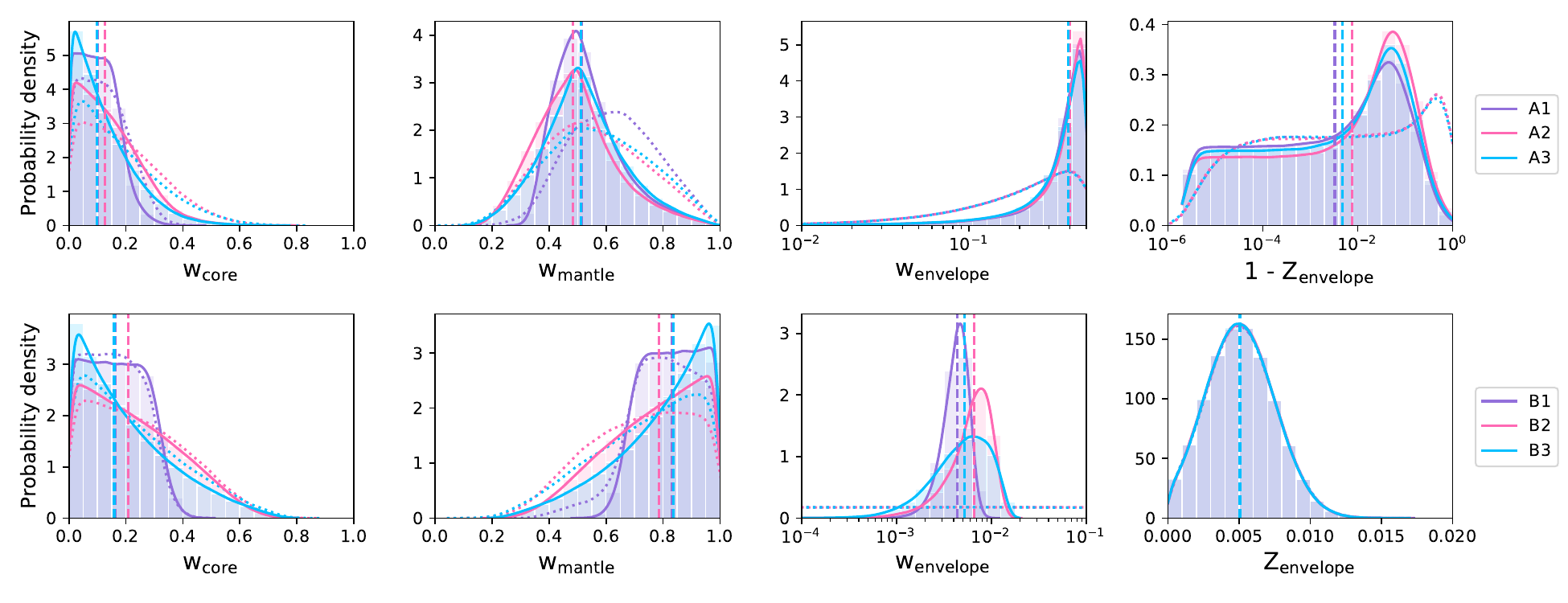}
    \caption{    
    Same as Fig.~\ref{fig:toi_5788b_interior} but for TOI-5788~c.
    }
    \label{fig:toi_5788c_interior}
\end{figure*}

With both mass and radius measurements of TOI-5788 b and c, we are able to make inferences as to the interior structure of the planets.
In Fig.~\ref{fig:mass_radius}, we place the TOI-5788 planets on a mass--radius diagram. 
We plot theoretical mass-radius curves of pure iron, Earth-like (32.5 per cent iron and 67.5 per cent MgSiO$_3$), and pure rocky (MgSiO$_3$) planets \citep{2019PNAS..116.9723Z} as well as the water-world and gas-dwarf (photoevaporation and core-assisted mass loss) models described by \citet{Rogers23}. 
The shaded regions indicate the agnostic initial conditions for the photoevaporation and core-powered mass loss scenarios, and the central models indicates boil-off initial conditions. 
We also overlay known exoplanets with both mass and radius measurements, and with radii $R\leq 6\,\mathrm{R_\oplus}$. 
To ensure high quality data, we require 
\begin{equation}
    \frac{\sigma_\rho}{\rho} = \frac{\sigma_M}{M} + 3\frac{\sigma_R}{R} \leq 0.5,
\end{equation}
where $\rho$ is the planet density.
This analysis indicates that, whilst TOI-5788~b is compatible with having a rocky or Earth-like composition, TOI-5788~c falls into a more degenerate region of parameter space, being consistent with both water-world and gas-dwarf models.

In an attempt to further probe the internal structure of both planets, we perform  internal-structure modelling using the \texttt{plaNETic} code\footnote{\url{https://github.com/joannegger/plaNETic}} \citep{Egger24}.
The neural network of \texttt{plaNETic} offers dramatic performance benefits over the forward model used in the classical Bayesian inference framework of \texttt{BICEPS} \citep{Haldemann24}.
The planetary model consists of three layers: an inner iron/sulphur core, a mantle of oxidised silicon, magnesium, and iron, and a volatile envelope comprising uniformly mixed hydrogen, helium, and water.
By default, the \texttt{plaNETic} framework computes an internal structure inference for three different priors of the planetary Si/Mg/Fe ratios for both water-rich and water-poor formation scenarios (i.e., formation from within and without the ice line, respectively), for a total of six models.
\edit{The models used are summarised in Table~\ref{tab:model_names}.}
In particular the \texttt{plaNETic} code considers the case where the Si:Mg:Fe ratio matches that of the host star \citep{Thiabaud15}, the case where the planet is iron-enriched relative to the star \citep{Adibekyan21}, and finally the case where the molar fractions of Si, Mg, and Fe are allowed to vary freely and are sampled uniformly from the simplex wherein the ratios sum to unity.
The priors for the Si:Mg:Fe ratio and water fraction are the same as were used to model the planets of the HIP~29442 system \citep{Egger24}, and are described in more detail in that work.

\begin{table}
    \centering
    \caption{\edit{A summary of the models used in the \texttt{plaNETic} analysis of TOI-5788~b and TOI-5788~c}}
    \begin{tabular}{c|c|cc}
    \hline \hline
         &  & Water-rich&Water-poor\\
         \hline
\multirow{3}{*}{\rotatebox[origin=c]{90}{Silicates}}& Stellar-like & A1&B1\\
 & Iron-rich& A2&B2\\
 & Free & A3&B3\\
 \hline \hline
 \end{tabular}
    \label{tab:model_names}
\end{table}

Figs.~\ref{fig:toi_5788b_interior} and~\ref{fig:toi_5788c_interior} show the results of our internal structure modelling\footnote{The full output for these models is shown in Appendix~\ref{sec:app_interior_models}.}. 
We find good agreement between the different silicate ratio models, indicating that our results are fairly robust to choice of silicate priors. 
For TOI-5788~b, the core and mantle mass ratios are all within 1-$\sigma$. 
This is because the envelope is small enough that the presence or absence of a significant water content does little to affect the bulk of the planet composition.
By contrast, for TOI-5788~c, the difference between the water-poor and water-rich models is more apparent. 
The water-rich models have a significantly larger envelope mass fraction (and correspondingly lower core and mantle fractions) than the water-poor models. 
This reflects the known degeneracy between the water-world and gas-dwarf models of sub-Neptunes \citep[see][and references therein]{Rogers23}.

\subsection{Atmospheric evolution}\label{sec:atmos_evolution}

\begin{figure}    
    \centering
    \includegraphics[width=1\linewidth]{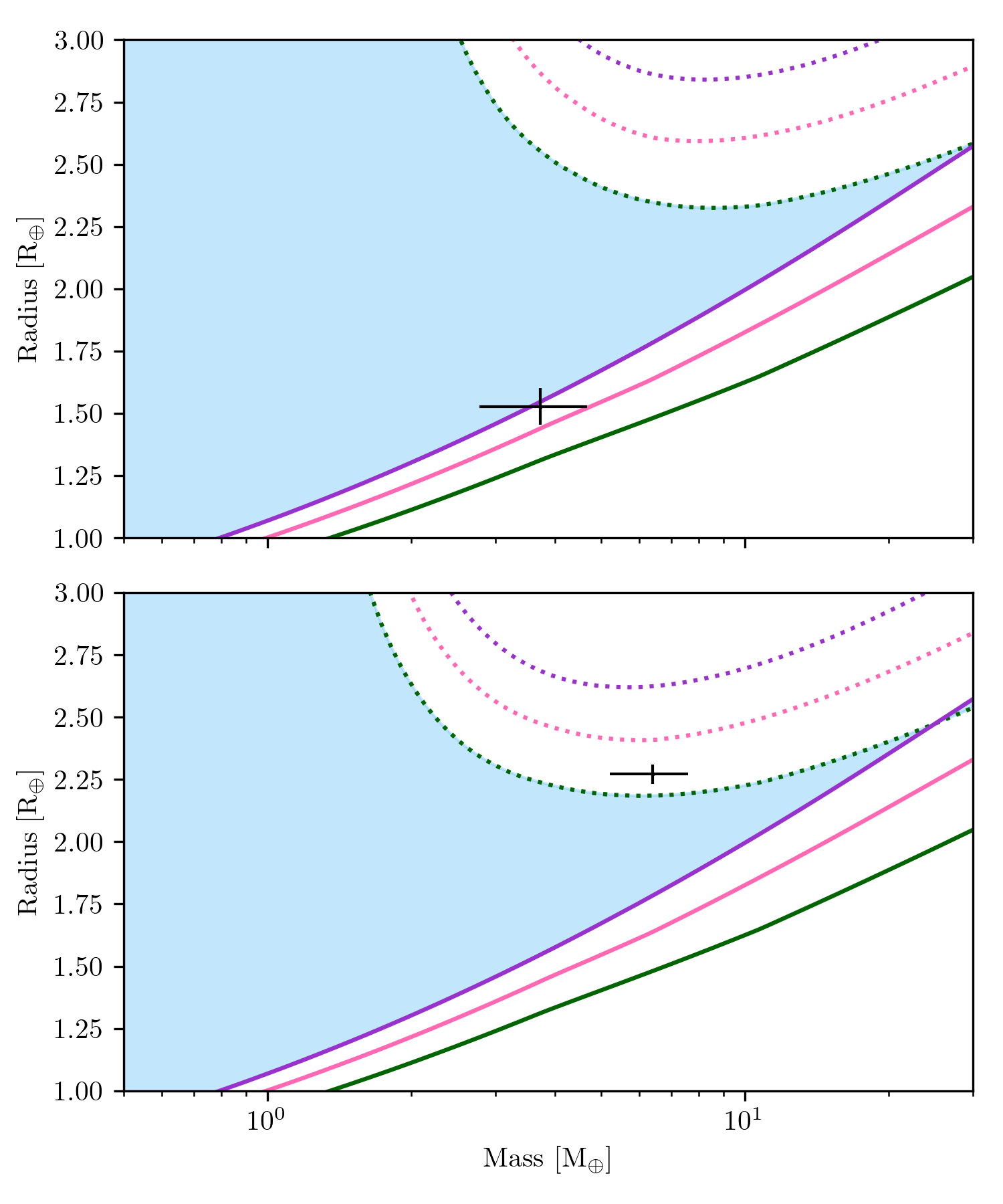}
    \caption{Mass--radius diagrams indicating the hot-water-world triangles of \citet{Egger2025A&A...696A..28E} as blue shaded regions. 
    This is the region of parameter space in which purely H/He atmospheres are expected to have evaporated within a few tens of Myr. 
    The upper and lower panels show the hot-water-world triangles for equilibrium temperatures of 970 \si{\kelvin} (corresponding to planet~b) and 707\,\si{\kelvin} (corresponding to planet~c), respectively.
    The green pink and purple lines correspond to models with Mercury-like cores, Earth-like cores, and pure silicate cores, respectively. 
    Solid lines are the relationships for the bare cores, and dotted lines show the relationships for planets with a 1 per cent H/He atmosphere.
    The masses and radii of planets b and c are shown in the top and bottom panels, respectively.}

    \label{fig:hot_water_world}
\end{figure}

We note that TOI-5788~b falls into an interesting region of parameter space of small, hot planets.
It is possible to lift some of the intrinsic degeneracy in compositions derived solely from mass and radius measurements by noting that purely H/He atmospheres  are not thought to be stable for small, highly irradiated planets \citep[e.g.,][]{2017ApJ...847...29O}. 
This would imply a significant fraction of the atmosphere of such planets must consist of heavier volatiles, the most common of which is water. 
\citet{Egger2025A&A...696A..28E} defined such regions in the mass-radius diagrams (Hot Water World, HWW, triangles) wherein pure H/He envelopes would not be stable, meaning that the atmospheres of planets falling into these regions need to contain at least some heavier volatiles
(see Fig.~\ref{fig:hot_water_world}). 
TOI-5788~b lies at the lower edge of the HWW triangle. 
Its position in a mass-radius diagram (Figs.~\ref{fig:mass_radius} and ~\ref{fig:hot_water_world}) shows that TOI-5788~b is consistent with a bare rocky core.  
\edit{Whilst such a formation pathway can produce the observed mass and radius, it poses difficulties from the point of view of formation models, as it implies that the planet accreted only rocky material and no iron at all.}
\edit{Its position above the Earth-like model implies that another plausible scenario for TOI-5788~b is an Earth-like core with some amount of atmosphere.}
\
TOI-5788~c lies in the region where the H/He atmospheres are expected to be stable, but nevertheless close above the HWW triangle.

To assess this scenario, we analysed the stability of H/He and 100 per cent water vapour atmospheres for both planets in the system. 
In the first case, we employed the atmospheric evolution models based on Modules for Experiments in Stellar Astrophysics \citep[MESA; see][]{paxton2013ApJS..208....4P} framework presented in \citet{kubyshkina2020MNRAS.499...77K} and \citet{Kubyshkina2022A&A...668A.178K}, combining the thermal evolution of H/He atmospheres with hydrodynamic atmospheric escape \citep{Reza2025A&A...694A..88R}.
For each planet, we probed the mass ranges reflecting the observational constraints.
\edit{Initial atmospheric mass fractions for the H/He atmosphere models, $w^0_{\rm envelope}\,\sim\,$0.5--10 per cent,  were taken from the value given by the analytical approximation by \citet{mordasini2020A&A...638A..52M}.
For the water vapour atmosphere models, $w^0_\text{envelope}$ is allowed to vary freely.}
We adopt the orbital distances provided in Table~\ref{tab:adopted_parameters}. 
For the stellar mass, we take a value of $0.9\mathrm{M_\odot}$, this is consistent with the value quoted in Table~\ref{sec:stellar_parameters} whilst minimising the differences in stellar temperatures with the stellar evolution model used in our simulations \citep{johnstone2021A&A...649A..96J}. 
We also considered different scenarios for the stellar rotation evolution of TOI-5788 as a proxy for the activity history of the star. 
This is quantified by the rotation period at a stellar age of 150 Myr, $P_\mathrm{rot}^{150}$.
We consider values of $P_\mathrm{rot}^{150}$ between 1 and 15 days. 
This covers a wide range of expected activity evolution history.
In total, we ran 48 models for TOI-5788\,b and 96 models for TOI-5788\,c.
We find an evaporation timescale of a H/He atmosphere around TOI-5788\,b to be between $\sim$0.01-200\,Myr. 
This implies that a H/He atmosphere would not be stable for TOI-5788~b at its current age of $\gtrsim 3.1$ Gyr.
On the other hand, TOI-5788\,c was reproduced for each considered stellar history, for the initial mass of the planet $\gtrsim$6.5\,$\mathrm{M_{\oplus}}$ and $w^0_{\rm envelope}\sim3$--25.5 per cent. 
The present-day atmospheric mass fractions predicted by the evolution model for these cases range between 0.1--1 per cent of the planet's mass, which is consistent with the analysis performed in Sec.\,\ref{sec:comp}. 
Therefore, retaining the primordial H/He-dominated atmosphere is a plausible scenario for planet c.

We also analysed the stability of water vapour in the planet atmospheres. 
Our approach is described in more detail in \citet{Egger2025A&A...696A..28E}, though we provide a brief summary here. 
For each planet, we run backwards models to probe the atmospheric evolution for a grid of planetary masses, radii, and system age ranging within the observational constraints and for all possible combinations of the following parameters: planetary cores of stellar, iron-rich, and unconstrained compositions (models 1-3 in Sec.\,\ref{sec:comp}, quantified here by the core silicate mass fractions $w_\mathrm{s}$), core luminosities $L_{\rm core}$ of $10^{19}$, $10^{21}$, and $10^{23}$\,erg\,s$^{-1}$ (kept constant throughout the evolution), and the atmospheric heating efficiency parameter $\eta$ of 1, 5, 10, and 15 per cent. 
The value of $\eta\sim$10-15 per cent is commonly considered plausible for H/He dominated atmospheres, whilst for water-rich atmospheres, $\eta$ is expected to be significantly lower \citep[e.g.][]{johnstone2020ApJ...890...79J}. 
The hydrodynamic simulations performed in \citet{Egger2025A&A...696A..28E} suggest that at the young ages (when the atmospheric escape is most relevant) $\eta\simeq5$ per cent is a reasonable estimation for hot mini-Neptunes similar to TOI\,5788\,b, whilst for the cooler TOI\,5788\,c $\eta$ is likely closer to 1 per cent \citep[see, e.g., estimates for GJ\,9827\,d][]{Piaulet2024ApJ...974L..10P}. 
For the stellar input, we use the same models as for H/He atmospheres and $P_{\rm rot}^{150}$ of 1 and 15 days.

\begin{figure}
    \centering
    \includegraphics[width=1.0\linewidth]{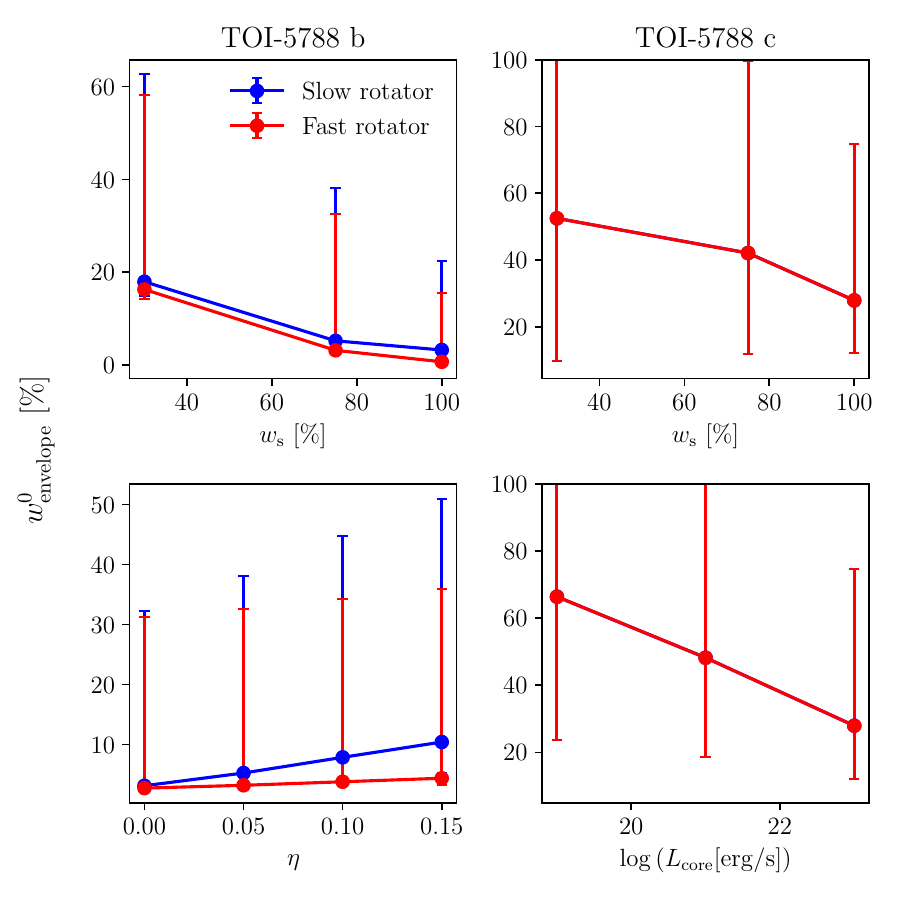}
    \caption{The values of $w^0_{\rm envelope}$ predicted by evolution models for water-atmospheres of TOI-5788\,b (left column) and TOI-5788\,c (right column) against the silicate mass fraction of the core ($w_\mathrm{s}$), atmospheric heating parameter ($\eta$), and core luminosity ($L_\mathrm{core}$). The parameters not quoted in each plot are kept at the nominal values. \edit{We plot the minimum, median, and maximum values of $w_\text{envelope}^0$}.}
    \label{fig:atm_evolution}
\end{figure}
In the top panels of Fig. \ref{fig:atm_evolution}, we show the relationship between $w^0_{\rm envelope}$ and the core composition (here quantified as the core silicates mass fraction $w_\mathrm{s}$). 
We assume the following nominal parameters for each planet: stellar-like composition core with $L_{\rm core} = 10^{21}$\,erg\,s$^{-1}$ and $\eta = 5$ per cent for TOI-5788~b and metal-poor core with $L_{\rm core} = 10^{23}$\,erg\,s$^{-1}$ and $\eta = 1$ per cent for TOI-5788~c.
For planet b, the water steam atmosphere is stable for most of the parameter space, and $w^0_{\rm envelope}$ values range from 4.6 to 33 per cent for the nominal conditions, which is consistent with the predictions of planetary formation models for similar planets.
This estimate depends only weakly on the exact present-day age of the system or the core luminosity. 
The strongest dependence is seen on the core composition where we obtain the largest value of $w^0_{\rm envelope}$ for the metal-rich core.
The dependence on $\eta$ is only significant in the case of the active star (see the lower right panel of Fig.\,\ref{fig:atm_evolution}).

For TOI-5788\,c, $w^0_{\rm envelope}$ varies between 16 and 47 per cent for the nominal parameters.
If, however, we consider the stellar composition core and moderate $L_{\rm core}$, as for planet b, the median value of $w^0_{\rm envelope}$ increases to $\sim$58 per cent and to 65 per cent for metal-rich core, which is hard to explain from the point of view of formation models.
Therefore, although we cannot fully rule out the scenario where TOI-5788\,c evolved with a water-dominated atmosphere, the H/He atmosphere scenario appears more realistic. 
It is worth noting that the gas dwarf model poses its own theoretical difficulties. 
\edit{\citet{2020A&A...644A.174V} show that, for planets formed via pebble accretion, dust growth is limited by fragmentation. 
A consequence of this is that the result of pebble-accretion planet formation is sensitive to the disc turbulence, with low ($\alpha \lesssim 10^{-4}$) viscosity discs typically producing terrestrial ($M \lesssim 3 M_\oplus$) cores.
\citet{2020A&A...644A.174V} also show that planets with core masses ($M_\mathrm{core} \approx 5 M_\oplus$) are able to retain significant H/He envelopes, resulting in gas-dwarf planets. 
Interestingly, they do not obtain any planets with radii between 2.5$R_\oplus$ and 6$R_\oplus$, showing that a combination of gas-dwarf and water-world evolution scenarios are needed to reproduce the observed planetary population.}

We also note that, as with the case of TOI-5788~b, the derived values of $w^0_{\rm envelope}$ vary only weakly with the age of the system.
Furthermore, the dependence on $\eta$ weakens even more and the difference between different stellar rotation scenarios (which is already relatively small for planet b) becomes negligible (lower right panel of Fig.~\ref{fig:atm_evolution}).
By contrast, the dependence on the core parameters remains strong.
As with TOI-5788~b, we find the largest value of $w^0_{\rm envelope}$ for the metal-rich core scenario.

It is important to note that for both planets, and both atmospheric evolution scenarios considered above, we see a remarkably weak dependence on the assumptions about the stellar evolution history, which is usually one of the largest uncertainties in atmospheric evolution studies \citep[e.g.][]{Kubyshkina2022A&A...668A.178K}. This makes the TOI-5788 system a promising target to test internal structure and atmospheric evolution models,
allowing us to better connect the present-day observations with the primordial state of the system.

\section{Prospects for follow-up observations}\label{sec:follow_up}

\subsection{Further RV observations}

Although we obtain a 5-$\sigma$ mass measurement of TOI-5788~c, we are unable to detect TOI-5788~b to the same level of precision.
To estimate the number of further observations needed to achieve a similar precision for TOI-5788~b, we follow the prescription of \citet{Cloutier18}, who show that, in the absence of correlated noise, the achievable RV precision from $N$ observations is 
\begin{equation} \label{eq:cloutier}
    \sigma_K = \sigma _\mathrm{RV} \sqrt{\frac{2}{N}}, 
\end{equation}
where $\sigma_\mathrm{RV}$ is the measurement uncertainty for each observation.
Although the derivation of \citet{Cloutier18} was for a single Keplerian signal in white noise, it can be shown that Eq.~\ref{eq:cloutier} holds for multi-planet systems, as long as the orbits of all planets are well sampled by the data.
To estimate $\sigma_\mathrm{RV}$, we identify a contribution from the instrument $\sigma_\mathrm{inst}$ and a stellar component $\sigma_\mathrm{st}$,
Taking fiducial values of $\sigma_\mathrm{inst} = 2.76\,\si{\metre\per\second}$, the median RV uncertainty, and $\sigma_\mathrm{st} = 0.42\,\si{\metre\per\second}$, the jitter of the RV time series, we obtain
\begin{equation}
    \sigma_\mathrm{RV} = \sqrt{\sigma_\mathrm{inst}^2 + \sigma^2_\mathrm{st}} = 2.79\,\si{\metre\per\second}.
\end{equation}
Aiming for a 5-$\sigma$ detection of TOI-5788~b, taking $K_\mathrm{b} = 1.42\,\si{\metre\per\second}$ from Table~\ref{tab:rv_parameters}, we estimate that a total of 191 30-minute exposures are required.
Owing to the large amount of HARPS-N time this would require to further refine the mass of the inner planet, we do not recommend further RV follow up at this stage. 
We nevertheless note the importance of long-term RV monitoring of planet-hosting stars to investigate the possible presence of large, outer planets in the system.

\subsection{Atmospheric follow up} \label{sec:atmos_followup}

We have shown in Section \ref{sec:comp} that TOI-5788~c occupies an interesting and highly-degenerate region of mass-radius space, being consistent with both a 50 per cent water world and a rocky planet with a 0.1-1 per cent H$_2$ atmosphere.
Although the atmospheric evolutionary analysis of Section~\ref{sec:atmos_evolution} favours the latter option, a model-independent way to break this degeneracy is via spectroscopic measurement of the planet's atmosphere.
In this Section, we will discuss the feasibility of such a follow up programme with the James Webb Space Telescope (\textit{JWST}).

\subsubsection{Transmission spectroscopy} \label{sec:tsm}

By observing starlight that has been attenuated by the atmosphere of an exoplanet, it is possible to infer the atmospheric chemical composition.
The ideal target for transmission spectroscopy is a hot, low density planet orbiting a small, bright star. 
To quantify the prospects of an individual target for transmission spectroscopy with \textit{JWST}, \citet{Kempton18} produce a Transmission Spectroscopy Metric (TSM). 
The TSM is defined as
\begin{equation}\label{eq:tsm}
    \mathrm{TSM} = \alpha 
    \left ( \frac{R_\mathrm{p}}{\mathrm{R_\oplus}}\right)^3
    \left ( \frac{M_\mathrm{p}}{\mathrm{M_\oplus}}\right)^{-1}
    \left ( \frac{R_{\star}}{\mathrm{R_\odot}}\right)^{-2}
    \left ( \frac{T_\mathrm{eq}}{\mathrm{K}}\right)
    \times 10^{-m_J/5},
\end{equation}
where $m_J$ is the $J$-band apparent magnitude, and $\alpha$ is a factor of order unity which varies with planet radius.
For TOI-5788~c, $\alpha = 1.26.$
It is important to note that, unlike the equilibrium temperature quoted in Section~\ref{sec:instellation}, Eq.~\ref{eq:tsm} assumes zero albedo.
Using our adopted parameters, we obtain a TSM of 36.5 for TOI-5788~c.

\subsubsection{Emission spectroscopy}

An alternative way of observing an exoplanet's atmosphere is to study the reflected light from the surface during secondary eclipse,
Again, for such an observing programme, the ideal candidate is a large, hot planet orbiting a small bright star.
As with the TSM, \citet{Kempton18} produce an Emission Spectroscopy Metric (ESM) to estimate the suitability of reliable emission spectroscopy observation.
The ESM is defined as 
\begin{equation}
    \mathrm{ESM} = 4.29\times 10^6 \times\frac{B_{7.5}(T_\mathrm{day})}{B_{7.5}(T_\mathrm{eff})} \left(\frac{R_\mathrm{p}}{R_\star}\right)^2 \times 10^{-m_K/5},
\end{equation}
where $B_{7.5}(T)$ is the Planck function for a temperature $T$, evaluated at $7.5\si{\micro\meter}$ and $m_K$ is the apparent $K$-band magnitude.
The day-side temperature, $T_\mathrm{day}$ is taken to be $1.1 \times T_\mathrm{eq}$, following \citet{Kempton18}. 
For TOI-5788~c, the ESM has a value of 1.91.

Although both the TSM and ESM for TOI-5788~c fall below the values for which \citet{Kempton18} favour follow-up observations (90 and 7.5, respectively), the TOI-5788 system offers an attractive opportunity to investigate planet formation around metal-poor stars.

\section{Conclusions}\label{sec:conclusions}

In this paper, we present the detection and characterisation of two transiting exoplanets orbiting the metal-poor solar-type star TOI-5788.
Using photometry from six \textit{TESS} sectors and three \textit{CHEOPS} visits, as well as radial velocities from a dedicated HARPS-N GTO campaign, we identify an inner super-Earth ($P = 6.340758\pm0.000030$\,d, $R = 1.528\pm0.075 \,\mathrm{R_\oplus}$) and outer mini-Neptune ($P = 16.213362\pm0.000026$\,d, $R=2.272\pm0.039\,\mathrm{R_\oplus}$).

From 125 HARPS-N observations, we report mass detections at the 4-$\sigma$ and 5-$\sigma$ level for TOI-5788~b ($3.72\pm0.94\,\mathrm{M_\oplus}$) and TOI-5788~c ($6.4\pm1.2\,\mathrm{M_\oplus}$), respectively.
We performed a stability analysis of the two-planet system and find potentially stable orbits between 8 and 14 days. 
Following this, we searched for a potential third planet in this stable region of parameter space but find no evidence of additional planets in the system.

We find that the TOI-5788 planets closely straddle the radius valley (Fig.~\ref{fig:radius_valley}).
The TOI-5788 system is one of very few metal-poor solar type stars with planets straddling the radius valley, making it an interesting system from the perspective of planet formation models. 

The precision to which we measure the masses and radii of planets~b and~c allows us to compare to the bulk-composition models of \citet{2019PNAS..116.9723Z, Rogers23}.
We show that TOI-5788~c falls in a highly degenerate region of mass-radius space, consistent with both a water-world model and a rocky core with a significant hydrogen envelope.
To break this degeneracy, we use \texttt{plaNETic} to model the interior structures of the planets (Figs.~\ref{fig:toi_5788b_interior}~\&~\ref{fig:toi_5788c_interior}). 
These models also allow for both a significant H/He atmosphere for TOI-5788~c as well as a water-world model. 
We also model the atmospheric evolution history of both planets. 
Whilst the water-world scenario for the formation and evolution of TOI-5788~c can be produced by the atmospheric evolutionary models of Section~\ref{sec:atmos_evolution}, \edit{our analysis hints that the gas-dwarf hypothesis is marginally preferred.}
For planet~b, our interior structure and atmospheric evolution models show that the planet is compatible with having a high mean molecular weight envelope, whereas a purely H/He envelope would not be stable to atmospheric mass loss.
Our interior modelling of TOI-5788~b, as well as the stellar abundances in Table~\ref{tab:abundances}, shows that TOI-5788~b follows the correlation between stellar and planetary iron composition reported by \citet{Adibekyan21}.

We also find that the results of the atmospheric evolution modelling are reasonably insensitive to assumptions about the stellar evolution history of TOI-5788.
Despite relatively low values of the TSM and ESM (Section~\ref{sec:atmos_followup}), and owing to the metal-poor nature of the star, the TOI-5788 system remains an interesting one for planetary atmospheric follow up.

\section*{Acknowledgements}

This work is based on observations made with the Italian Telescopio Nazionale Galileo (TNG) operated on the island of La Palma by the Fundaci\'on Galileo Galilei of the INAF (Instituto Nazionale di Astrofisica) at the Spanish Observatorio del Roque de los Muchachos of the Instituto de Astrofisica de Canarias. The HARPS-N project was funded by the Prodex Program of the Swiss Space Office (SSO), the Harvard University Origin of Life Initiative (HUOLI), the Scottish Universities Physics Alliance (SUPA), the University of Geneva, the Smithsonian Astrophysical Observatory (SAO), the Italian National Astrophysical Institute (INAF), University of St. Andrews, Queen’s University Belfast, and University of Edinburgh.

{\it CHEOPS} is an ESA mission in partnership with Switzerland with important contributions to the payload and the ground segment from Austria, Belgium, France, Germany, Hungary, Italy, Portugal, Spain, Sweden, and the United Kingdom. The {\it CHEOPS} Consortium would like to gratefully acknowledge the support received by all the agencies, offices, universities, and industries involved. Their flexibility and willingness to explore new approaches were essential to the success of this mission. {\it CHEOPS} data analysed in this article will be made available in the {\it CHEOPS} mission archive (\url{https://cheops.unige.ch/archive_browser/}).
The Belgian participation to CHEOPS has been supported by the Belgian Federal Science Policy Office (BELSPO) in the framework of the PRODEX Program, and by the University of Liege through an ARC grant for Concerted Research Actions financed by the Wallonia-Brussels Federation. MG is F.R.S.-FNRS Research Director.

This paper made use of data collected by the \textit{TESS} mission and are publicly available from the Mikulski Archive for Space Telescopes (MAST) operated by the Space Telescope Science Institute (STScI). Funding for the \textit{TESS} mission is provided by NASA’s Science Mission Directorate. We acknowledge the use of public \textit{TESS} data from pipelines at the \textit{TESS} Science Office and at the \textit{TESS} Science Processing Operations Center. Resources supporting this work were provided by the NASA High-End Computing (HEC) Program through the NASA Advanced Supercomputing (NAS) Division at Ames Research Center for the production of the SPOC data products. This research has made use of the Exoplanet Follow-up Observation Programme (ExoFOP; DOI: 10.26134/ExoFOP5) website, which is operated by the California Institute of Technology, under contract with the National Aeronautics and Space Administration under the Exoplanet Exploration Programme.

This work has made use of data from the European Space Agency (ESA) mission {\it Gaia} (\url{https://www.cosmos.esa.int/gaia}), processed by the {\it Gaia} Data Processing and Analysis Consortium (DPAC, \url{https://www.cosmos.esa.int/web/gaia/dpac/consortium}). Funding for the DPAC has been provided by national institutions, in particular the institutions participating in the {\it Gaia} Multilateral Agreement.

This publication makes use of The Data \& Analysis Center for Exoplanets (DACE), which is a facility based at the University of Geneva (CH) dedicated to extrasolar planets data visualisation, exchange and analysis. DACE is a platform of the Swiss National Centre of Competence in Research (NCCR) PlanetS, federating the Swiss expertise in Exoplanet research. The DACE platform is available at https://dace.unige.ch.

This work has been carried out within the framework of the NCCR PlanetS supported by the Swiss National Science Foundation under grants 51NF40\_182901 and 51NF40\_205606.
We acknowledge financial support from the Agencia Estatal de Investigaci\'on of the Ministerio de Ciencia e Innovaci\'on MCIN/AEI/10.13039/501100011033 and the ERDF “A way of making Europe” through project PID2021-125627OB-C32, and from the Centre of Excellence “Severo Ochoa” award to the Instituto de Astrofisica de Canarias.

BSL and AM acknowledge funding from a UKRI Future Leader Fellowship, grant number MR/X033244/1. AM acknowledges funding from a UK Science and Technology Facilities Council (STFC) small grant ST/Y002334/1. 
DAT acknowledges the support of the Science and Technology Facilities Council (STFC).
R.D.H. is funded by the UK Science and Technology Facilities Council (STFC)'s Ernest Rutherford Fellowship (grant number ST/V004735/1).
ZG acknowledges support from the ESA PRODEX projects Nos. 4000137122, 4000149202, and 4000149203 between ELTE Eötvös Loránd University and the
European Space Agency, as well as from the VEGA grant of the Slovak Academy of Sciences (No. 2/0033/26), the Slovak Research and Development Agency contract (No. APVV-24-0160), the support from SNN-147362 and the ADVANCED-153410 of the National Research, Development and Innovation Office (NKFIH, Hungary), and the support of the city of Szombathely.
This work has been carried out within the framework of the NCCR PlanetS supported by the Swiss National Science Foundation under grants 51NF40\_182901 and 51NF40\_205606.
JAE acknowledges support through the European Space Agency (ESA) Research Fellowship Programme in Space Science.
DK was supported by a Schr\"odinger Fellowship supported by the Austrian Science Fund (FWF) project number J4792 (FEPLowS).
CMC acknowledges support from the FCT, Portugal, through the CFisUC projects UIDB/04564/2020 770 and UIDP/04564/2020, with DOI identifiers 10.54499/UIDB/04564/2020 and 10.54499/UIDP/04564/2020, respectively.
PM acknowledges support from STFC research grant number ST/R000638/1. 
ML acknowledges support of the Swiss National Science Foundation under grant number PCEFP2\_194576. The contribution of ML has been carried out within the framework of the NCCR PlanetS supported by the Swiss National Science Foundation under grants 51NF40\_182901 and 51NF40\_205606.
RA acknowledges financial support from the Agencia Estatal de Investigación of the Ministerio de Ciencia e Innovación MCIN/AEI/10.13039/501100011033 and the ERDF “A way of making Europe” through projects PID2021-125627OB-C31, PID2021-125627OB-C32, PID2021-127289NB-I00, PID2023-150468NB-I00 and PID2023-149439NB-C41; from the Centre of Excellence “Severo Ochoa'' award to the Instituto de Astrofísica de Canarias (CEX2019-000920-S), the Centre of Excellence “María de Maeztu” award to the Institut de Ciències de l’Espai (CEX2020-001058-M), and from the Generalitat de Catalunya/CERCA programme.
NCSa acknowledges funding by the European Union (ERC, FIERCE, 101052347). Views and opinions expressed are however those of the author(s) only and do not necessarily reflect those of the European Union or the European Research Council. Neither the European Union nor the granting authority can be held responsible for them.
DG gratefully acknowledges financial support from the CRT foundation under Grant No. 2018.2323 “Gaseousor rocky? Unveiling the nature of small worlds”.
PRODEX Experiment Agreement No. 4000137122
CHEOPS is an ESA mission in partnership with Switzerland with important contributions to the payload and the ground segment from Austria, Belgium, France, Germany, Hungary, Italy, Portugal, Spain, Sweden, and the United Kingdom. The CHEOPS Consortium would like to gratefully acknowledge the support received by all the agencies, offices, universities, and industries involved. Their flexibility and willingness to explore new approaches were essential to the success of this mission. CHEOPS data analysed in this article will be made available in the CHEOPS mission archive (\url{https://cheops.unige.ch/archive\_browser}). This project has received funding from the Swiss National Science Foundation for project 200021\_200726. It has also been carried out within the framework of the National Centre of Competence in Research PlanetS supported by the Swiss National Science Foundation under grant 51NF40\_205606. The authors acknowledge the financial support of the SNSF.
X.D acknowledges the support from the European Research Council (ERC) under the European Union’s Horizon 2020 research and innovation programme (grant agreement SCORE No 851555) and from the Swiss National Science Foundation under the grant SPECTRE (No 200021\_215200)
C.A.W. would like to acknowledge support from the UK Science and Technology Facilities Council (STFC, grant number ST/X00094X/1).
This work was funded by the European Union (ERC, FIERCE, 101052347). Views and opinions expressed are however those of the author(s) only and do not necessarily reflect those of the European Union or the European Research Council. Neither the European Union nor the granting authority can be held responsible for them. This work was also supported by FCT - Fundação para a Ciência e a Tecnologia through national funds by these grants: UIDB/04434/2020 DOI: 10.54499/UIDB/04434/2020, UIDP/04434/2020 DOI: 10.54499/UIDP/04434/2020, PTDC/FIS-AST/4862/2020, UID/04434/2025.
This project has been carried out within the framework of the National Centre of Competence in Research PlanetS supported by the Swiss National Science Foundation under grant 51NF40\_205606. The authors acknowledge the financial support of the SNSF.
M.P. acknowledges support from the European Union – NextGenerationEU (PRIN MUR 2022 20229R43BH), the ``Programma di Ricerca Fondamentale INAF 2023'', and from the Italian Space Agency (ASI) under contract 2018-24-HH.0 ``The Italian participation in the Gaia Data Processing and Analysis Consortium (DPAC)'' in collaboration with the Italian National Institute of Astrophysics.
T.G.W acknowledges support from the University of Warwick and UKSA.
S.M.O. is supported by a UK Science and Technology Facilities Council (STFC) Studentship (ST/W507751/1). S.M.O. acknowledges the support of the Royal Astronomical Society in the form of an RAS Observing/Field Trip Travel Grant.
This work was funded by the European Union (ERC, FIERCE, 101052347). Views and opinions expressed are however those of the author(s) only and do not necessarily reflect those of the European Union or the European Research Council. Neither the European Union nor the granting authority can be held responsible for them. This work was also supported by FCT - Fundação para a Ciência e a Tecnologia through national funds by these grants: UIDB/04434/2020 DOI: 10.54499/UIDB/04434/2020, UIDP/04434/2020 DOI: 10.54499/UIDP/04434/2020, PTDC/FIS-AST/4862/2020, UID/04434/2025.
M.S. acknowledges financial support from the Belgian Federal Science Policy Office (BELSPO) in the framework of the PRODEX Programme of the European Space Agency (ESA) under contract number C4000140754.
CH’s contribution is supported by NASA under award number 80GSFC24M0006.
DR was supported by NASA under award number 80NSSC25M7110.

\section*{Data Availability}

The raw and detrended photometric CHEOPS time series data, as well as the radial velocity measurements, will be made available in electronic form on CDS upon publication.



\bibliographystyle{mnras}
\bibliography{example} 




\appendix

\section{Full results from interior modelling}  \label{sec:app_interior_models}

In Tables~\ref{tab:internal_structure_results_b} and~\ref{tab:internal_structure_results_c} we present the full results from the interior modelling of TOI-5788~b and TOI-5788~c, repesctively. 
See Section\ref{sec:comp} for description of the models.

\begin{table*}
\renewcommand{\arraystretch}{1.5}
\caption{Results of the internal structure modelling for TOI-5788 b.}
\centering
\begin{tabular}{r|ccc|ccc}
\hline \hline
Water prior &              \multicolumn{3}{c|}{Water-rich prior (formation outside iceline)} & \multicolumn{3}{c}{Water-poor prior (formation inside iceline)} \\
Si/Mg/Fe prior &           Stellar (A1) &       Iron-enriched (A2) &      Free (A3) &
                           Stellar (B1) &       Iron-enriched (B2) &      Free (B3) \\
\hline
w$_\textrm{core}$ [\%] &        $16_{-11}^{+10}$ &    $24_{-16}^{+19}$ &    $24_{-17}^{+21}$ &
                           $18_{-12}^{+11}$ &    $29_{-19}^{+20}$ &    $31_{-21}^{+22}$ \\
w$_\textrm{mantle}$ [\%] &      $77_{-11}^{+12}$ &    $68_{-19}^{+17}$ &    $68_{-22}^{+18}$ &
                           $82_{-11}^{+12}$ &    $71_{-20}^{+19}$ &    $69_{-22}^{+21}$ \\
w$_\textrm{envelope}$ [\%] &    $4.7_{-3.4}^{+7.2}$ &    $6.0_{-4.1}^{+7.5}$ &    $6.3_{-4.3}^{+7.7}$ &
                           $\left(1.7_{-0.5}^{+1.5}\right)$ $10^{-4}$ &    $\left(2.8_{-1.4}^{+5.3}\right)$ $10^{-4}$ &    $\left(3.3_{-1.9}^{+8.0}\right)$ $10^{-4}$ \\
\hline
Z$_\textrm{envelope}$ [\%] &        $99.9_{-1.2}^{+0.1}$ &    $99.9_{-1.4}^{+0.1}$ &    $99.9_{-1.4}^{+0.1}$ &
                           $0.5_{-0.2}^{+0.2}$ &    $0.5_{-0.2}^{+0.2}$ &    $0.5_{-0.2}^{+0.2}$ \\
\hline
x$_\textrm{Fe,core}$ [\%] &     $90.3_{-6.4}^{+6.6}$ &    $90.4_{-6.4}^{+6.5}$ &    $90.4_{-6.4}^{+6.5}$ &
                           $90.2_{-6.3}^{+6.6}$ &    $90.3_{-6.4}^{+6.5}$ &    $90.4_{-6.4}^{+6.5}$ \\
x$_\textrm{S,core}$ [\%] &      $9.7_{-6.6}^{+6.4}$ &    $9.6_{-6.5}^{+6.4}$ &    $9.6_{-6.5}^{+6.4}$ &
                           $9.8_{-6.6}^{+6.3}$ &    $9.7_{-6.5}^{+6.4}$ &    $9.6_{-6.5}^{+6.4}$ \\
\hline
x$_\textrm{Si,mantle}$ [\%] &   $42_{-8}^{+10}$ &    $35_{-9}^{+12}$ &    $28_{-19}^{+29}$ &
                           $43_{-8}^{+10}$ &    $34_{-9}^{+12}$ &    $25_{-18}^{+28}$ \\
x$_\textrm{Mg,mantle}$ [\%] &   $41_{-10}^{+9}$ &    $34_{-12}^{+12}$ &    $34_{-22}^{+26}$ &
                           $41_{-11}^{+9}$ &    $33_{-11}^{+12}$ &    $34_{-21}^{+27}$ \\
x$_\textrm{Fe,mantle}$ [\%] &   $16_{-10}^{+9}$ &    $30_{-19}^{+19}$ &    $31_{-21}^{+23}$ &
                           $16_{-10}^{+10}$ &    $32_{-20}^{+18}$ &    $34_{-22}^{+22}$ \\
\hline
\end{tabular}
\label{tab:internal_structure_results_b}
\end{table*}
\renewcommand{\arraystretch}{1.0}

\begin{table*}
\renewcommand{\arraystretch}{1.5}
\caption{Results of the internal structure modelling for TOI-5788 c.}
\centering
\begin{tabular}{r|ccc|ccc}
\hline \hline
Water prior &              \multicolumn{3}{c|}{Water-rich prior (formation outside iceline)} & \multicolumn{3}{c}{Water-poor prior (formation inside iceline)} \\
Si/Mg/Fe prior &           Stellar (A1) &       Iron-enriched (A2) &      Free (A3) &
                           Stellar (B1) &       Iron-enriched (B2) &      Free (B3) \\
\hline
w$_\textrm{core}$ [\%] &        $10_{-7}^{+7}$ &    $13_{-9}^{+13}$ &    $10_{-7}^{+13}$ &
                           $16_{-11}^{+11}$ &    $21_{-15}^{+20}$ &    $16_{-12}^{+21}$ \\
w$_\textrm{mantle}$ [\%] &      $51_{-9}^{+13}$ &    $48_{-13}^{+14}$ &    $51_{-13}^{+15}$ &
                           $83_{-11}^{+11}$ &    $79_{-21}^{+15}$ &    $84_{-22}^{+12}$ \\
w$_\textrm{envelope}$ [\%] &    $39_{-15}^{+8}$ &    $40_{-17}^{+7}$ &    $39_{-16}^{+8}$ &
                           $0.43_{-0.13}^{+0.13}$ &    $0.66_{-0.30}^{+0.31}$ &    $0.52_{-0.29}^{+0.43}$ \\
\hline
Z$_\textrm{envelope}$ [\%] &        $99.7_{-6.6}^{+0.3}$ &    $99.2_{-8.6}^{+0.8}$ &    $99.5_{-7.5}^{+0.5}$ &
                           $0.5_{-0.2}^{+0.2}$ &    $0.5_{-0.2}^{+0.2}$ &    $0.5_{-0.2}^{+0.2}$ \\
\hline
x$_\textrm{Fe,core}$ [\%] &     $90.3_{-6.4}^{+6.5}$ &    $90.3_{-6.4}^{+6.5}$ &    $90.3_{-6.4}^{+6.5}$ &
                           $90.3_{-6.4}^{+6.6}$ &    $90.4_{-6.4}^{+6.5}$ &    $90.4_{-6.4}^{+6.5}$ \\
x$_\textrm{S,core}$ [\%] &      $9.7_{-6.5}^{+6.4}$ &    $9.7_{-6.5}^{+6.4}$ &    $9.7_{-6.5}^{+6.4}$ &
                           $9.7_{-6.6}^{+6.4}$ &    $9.6_{-6.5}^{+6.4}$ &    $9.6_{-6.5}^{+6.4}$ \\
\hline
x$_\textrm{Si,mantle}$ [\%] &   $42_{-7}^{+9}$ &    $38_{-10}^{+11}$ &    $37_{-26}^{+29}$ &
                           $42_{-8}^{+10}$ &    $37_{-10}^{+11}$ &    $36_{-25}^{+29}$ \\
x$_\textrm{Mg,mantle}$ [\%] &   $41_{-11}^{+9}$ &    $36_{-13}^{+12}$ &    $37_{-26}^{+32}$ &
                           $40_{-11}^{+10}$ &    $36_{-13}^{+12}$ &    $37_{-25}^{+30}$ \\
x$_\textrm{Fe,mantle}$ [\%] &   $17_{-11}^{+9}$ &    $25_{-17}^{+20}$ &    $18_{-13}^{+23}$ &
                           $17_{-11}^{+9}$ &    $25_{-17}^{+20}$ &    $18_{-14}^{+24}$ \\
\hline
\end{tabular}
\label{tab:internal_structure_results_c}
\end{table*}
\renewcommand{\arraystretch}{1.0}


\bsp	
\label{lastpage}
\end{document}